\documentclass[a4paper,11pt]{article}
\pdfoutput=1

\usepackage{jcappub}
\usepackage[utf8]{inputenc}
\usepackage{amsmath,mathtools}
\usepackage{tensor}
\usepackage{amssymb}
\usepackage{graphicx,epsfig,amssymb}
\usepackage{physics}
\usepackage{lipsum}  
\usepackage{bigints}
\usepackage{verbatim}
\usepackage{hyperref}
\usepackage{color}
\usepackage{multirow}
\usepackage{float}
\usepackage{array}
\usepackage{orcidlink}
\usepackage{booktabs}

\usepackage{booktabs}
\usepackage{array}
\usepackage{ragged2e}
\usepackage{makecell}

\newcolumntype{P}[1]{>{\centering\arraybackslash}p{#1}}

\newcommand{\ede}{_{\rm EDE}}

\begin{document}

\title{Unifying Early and Late Dark Energy:~Dynamical Requirements and Obstructions}

\author[a]{William Giar\`{e}}
\author[a]{Jeremy Sakstein}
\affiliation[a]{Department of Physics and Astronomy, University of Hawai‘i, Honolulu, HI 96822, USA}
\emailAdd{giare@hawaii.edu}
\emailAdd{sakstein@hawaii.edu}

\abstract{We investigate whether early- and late-time dark energy could arise from a single scalar field.~Adopting a bottom-up perspective, we first identify the sequence of dynamical regimes that any unified scenario must traverse to account for both an early dark energy phase relevant for pre-recombination solutions of the Hubble tension and the late-time acceleration of the Universe.~We derive the corresponding requirements on the scalar energy density and equation of state.~We then adopt a complementary top-down perspective and translate these requirements into constraints on the phase-space structure of minimally coupled scalar fields with tracking-like dynamics.~We show that satisfying all requirements necessitates a potential with three distinct slopes, arranged in a steep-steeper–shallow hierarchy.~This conclusion remains unchanged in the presence of conformal couplings to dark matter.~These results place strong constraints on attempts to construct unified models of early- and late-time dark energy.~We discuss implications for model-building. }

\maketitle

\section{Introduction}
\label{sec:introduction}

As first indicated by observations of distant Type Ia supernovae (SN)~\cite{SupernovaSearchTeam:1998fmf,SupernovaCosmologyProject:1998vns}, and subsequently corroborated by a wide range of independent cosmological probes~\cite{Sherwin:2011gv,Moresco:2016mzx,Rubin:2016iqe,Haridasu:2017lma,Planck:2018vyg,Yang:2019fjt,DiValentino:2020evt,Nadathur:2020kvq,Rose:2020shp,DES:2020cbm,eBOSS:2020yzd,ACT:2023dou,ACT:2023kun,SPT-3G:2024atg,DESI:2024mwx,Reischke:2025hrt,SPT-3G:2025bzu,Stolzner:2025htz,Wright:2025xka,DESI:2025zgx,DES:2026fyc,DES:2026qfi}, the Universe is currently undergoing a phase of accelerated expansion.~In General Relativity (GR), such behavior requires a component of Dark Energy (DE), whose effective pressure is sufficiently negative to overcome the decelerating effect of (dark) matter and radiation.~The standard $\Lambda$ Cold Dark Matter ($\Lambda$CDM) cosmological model attributes DE to a cosmological constant $\Lambda$.~{However,} recent DESI baryon acoustic oscillation (BAO) measurements~\cite{DESI:2024mwx,DESI:2025zgx}, when combined with Cosmic Microwave Background (CMB)~\cite{Planck:2018vyg,ACT:2023dou,ACT:2023kun,SPT-3G:2024atg,SPT-3G:2025bzu} and SN~\cite{Scolnic:2021amr,DES:2025sig,Hoyt:2026fve} data, exhibit a preference for an evolving DE equation of state~\cite{DESI:2025fii}.~{If confirmed, this would point beyond a purely constant vacuum energy,} suggesting
the possible presence of dynamical degrees of freedom, such as scalar fields.

A separate challenge for the $\Lambda$CDM framework is the Hubble tension~\cite{Verde:2019ivm,DiValentino:2020zio,DiValentino:2021izs,Perivolaropoulos:2021jda,Schoneberg:2021qvd,Shah:2021onj,Abdalla:2022yfr,DiValentino:2022fjm,Kamionkowski:2022pkx,Giare:2023xoc,Hu:2023jqc,Verde:2023lmm,CosmoVerseNetwork:2025alb}.~Local measurements of $H_0$, such as those reported by the SH0ES collaboration based on Type Ia SN ($H_0 = 73 \pm 1$ km/s/Mpc)~\cite{Riess:2021jrx}, are in $\gtrsim 5\sigma$ tension with values inferred from the CMB and other early-Universe probes within the $\Lambda$CDM framework, such as those obtained by the Planck collaboration ($H_0 = 67.4 \pm 0.5$ km/s/Mpc)~\cite{Planck:2018vyg}.~While unresolved systematics remain an actively investigated possibility~\cite{Efstathiou:2020wxn,Mortsell:2021nzg,Mortsell:2021tcx,Riess:2021jrx,Sharon:2023ioz,Murakami:2023xuy,Riess:2023bfx,Bhardwaj:2023mau,Brout:2023wol,Dwomoh:2023bro,Uddin:2023iob,Riess:2024ohe,Freedman:2024eph,Riess:2024vfa}, the remarkable internal consistency of early-Universe determinations around the Planck $\Lambda$CDM value, together with the coherent picture emerging from the local distance network~\cite{H0DN:2025lyy}, makes a purely systematic explanation increasingly difficult to sustain.~This, in turn, strengthens the case for exploring possible forms of new physics beyond the standard $\Lambda$CDM model~\cite{DiValentino:2021izs,Schoneberg:2021qvd,CosmoVerseNetwork:2025alb}.

Proposed solutions typically fall into two classes:~those that modify the expansion history before recombination (early-time solutions) and those that modify it after recombination (late-time solutions).~Early-time scenarios increase the expansion rate prior to recombination, reducing the sound horizon $r_s(z_\star)$ and thereby allowing for a larger inferred value of $H_0$ while preserving the observed angular scale $\theta_\star$ of the acoustic peaks.~{A prominent class of examples is provided by Early Dark Energy (EDE) models , which introduce a transient pre-recombination energy component.}\footnote{{The idea of EDE cosmologies dates back at least to Refs.~\cite{Wetterich:2004pv,Doran:2006kp}.~The literature on EDE models in relation to the Hubble tension is by now extensive.~For recent discussions spanning model-building aspects, phenomenological analyses, and updates on the status of the tension, we refer the reader to Refs.~\cite{Poulin:2018cxd,Alexander:2019rsc,Smith:2019ihp,Capparelli:2019rtn,Sakstein:2019fmf,Ye:2020btb,Chudaykin:2020acu,Gogoi:2020qif,Braglia:2020bym,Niedermann:2020dwg,Ivanov:2020ril,DAmico:2020ods,Weiner:2020sxn,Lin:2020jcb,Smith:2020rxx,Murgia:2020ryi,Jedamzik:2020zmd,Chudaykin:2020igl,Fujita:2020ecn,Adi:2020qqf,Seto:2021xua,Tian:2021omz,Freese:2021rjq,DiValentino:2021izs,Sabla:2021nfy,Nojiri:2021dze,Ye:2021nej,Boylan-Kolchin:2021fvy,Vagnozzi:2021gjh,Karwal:2021vpk,Jiang:2021bab,Schoneberg:2021qvd,Poulin:2021bjr,Niedermann:2021ijp,McDonough:2021pdg,LaPosta:2021pgm,Herold:2021ksg,Sabla:2022xzj,Ye:2022afu,Jiang:2022uyg,Rudelius:2022gyu,Berghaus:2022cwf,Ye:2022efx,Reeves:2022aoi,Alexander:2022own,Gomez-Valent:2022bku,Simon:2022adh,Rezazadeh:2022lsf,Smith:2022iax,McDonough:2022pku,Murai:2022zur,Wang:2022bmk,Forconi:2023hsj,Hart:2022agu,Secco:2022kqg,Herold:2022iib,Kamionkowski:2022pkx,Lin:2022phm,Brissenden:2023yko,Reboucas:2023rjm,Cruz:2023cxy,Poulin:2023lkg,Goldstein:2023gnw,Cicoli:2023qri,Jiang:2023bsz,Eskilt:2023nxm,Nojiri:2023mvi,Cruz:2023lmn,Cruz:2023lnq,Liu:2023kce,Peng:2023bik,FrancoAbellan:2023gec,Smith:2023oop,Ramadan:2023ivw,Fu:2023tfo,McDonough:2023qcu,Efstathiou:2023fbn,Gsponer:2023wpm,Khalife:2023qbu,Garny:2024ums,Giare:2024akf,Qu:2024lpx,Seto:2024cgo,Shen:2024hpx,Chatrchyan:2024xjj,Wang:2024tjd,SPT-3G:2025vyw,Carloni:2025jlk,Wang:2025dtk,Andriot:2025los,Poulin:2025nfb,Pang:2025lvh,Allali:2025wwi,Dialektopoulos:2025mfz,Jiang:2025hco,Liu:2026zqc,Bisabr:2026jeo,Yuan:2026xcg,Jhaveri:2026bla,Bella:2026zuk}.}}

{Late-time solutions instead assume a $\Lambda$CDM pre-recombination history and modify the post-recombination expansion to raise $H_0$.~While such scenarios face general difficulties that prevent them from fully resolving the tension on their own~\cite{Knox:2019rjx,Krishnan:2021dyb,Cai:2021weh,Cai:2022dkh,Raveri:2023zmr,Poulin:2024ken,Teixeira:2025czm,Pedrotti:2025ccw,Bansal:2026axl,Pedrotti:2026dwj},\footnote{{The main difficulty is that leaving the pre-recombination physics unchanged anchors the sound horizon to its $\Lambda$CDM value, making it challenging to fully reconcile the inverse-distance ladder with the local distance ladder~\cite{Raveri:2023zmr,Poulin:2024ken,Teixeira:2025czm}.}} they remain of considerable interest, especially in light of the recent DESI results.~The simplest dynamical DE models favored by DESI do not generally alleviate the Hubble tension~\cite{DESI:2024mwx,DESI:2025zgx}.~Alternative scenarios, such as interacting Dark Energy (IDE), can instead increase the inferred value of $H_0$ and preserve a good fit to both low- and high-redshift data~\cite{Giare:2024smz}.\footnote{{IDE cosmology has also a long history of study.~Earlier works have investigated its implications for the Hubble tension, while more recent studies have explored its connection to the DESI-preferred late-time dynamics.~For discussions spanning both directions, see Refs.~\cite{Amendola:2006dg,Wang:2010su,Baldi:2011th,Baldi:2011wy,Xia:2016vnp,DAmico:2016ntq,Guo:2017hea,DiValentino:2017iww,Zheng:2017asg,Yang:2017ccc,Yang:2018ubt,Yang:2018euj,Wang:2018azy,Yang:2018xlt,Wang:2018duq,Yang:2018uae,Li:2019loh,DiValentino:2019ffd,Feng:2019jqa,DiValentino:2019jae,Pan:2020zza,Pan:2020bur,Aljaf:2020eqh,Li:2020gtk,DiValentino:2020leo,DiValentino:2020vnx,DiValentino:2020kpf,Yang:2021hxg,Gao:2021xnk,Wang:2021kxc,Elizalde:2021kmo,Lucca:2021eqy,Nunes:2021zzi,Harko:2022unn,Yao:2022kub,Carrilho:2022mon,Califano:2022syd,Yang:2022csz,Ong:2022wrs,Gao:2022ahg,Zhai:2023yny,Bernui:2023byc,vanderWesthuizen:2023hcl,Hoerning:2023hks,Teixeira:2023zjt,Han:2023exn,Pan:2023mie,Benisty:2024lmj,Giare:2024ytc,Giare:2024smz,Sabogal:2024yha,Lewis:2024cqj,Li:2024qso,Carrion:2025bbk,Zhang:2025lam,Zhang:2025dwu,Yashiki:2025loj,Silva:2025hxw,Tsedrik:2025cwc,Sabogal:2025mkp,Li:2025owk,Figueruelo:2026eis,Escobal:2026zxb,Li:2026xaz,Dai:2026pvx,Kashyap:2026ivg,Kolhatkar:2026ixl,Antusch:2026ldp}}.}} 

Jointly accounting for the Hubble tension and the DESI preference for evolving dark energy raises the possibility that early- and late-time dark energy phenomena may be different manifestations of a common scalar dynamics.~Various model-specific realizations of this idea have been proposed~\cite{Nojiri:2021dze,Adil:2022hkj,Guendelman:2022cop,Copeland:2023zqz,Ramadan:2023ivw,Odintsov:2023cli,Brissenden:2023yko,Chowdhury:2023opo,Sohail:2024oki,Guendelman:2025swp,Sohail:2025mma}.~However, these constructions leave open the broader question of what dynamical structure is required for a unified early- and late-time dark energy scenario to be viable.

In this work, we address this question in two steps.~First, we adopt a bottom-up perspective and identify \textit{necessary} dynamical conditions that a scalar must satisfy to realize a unified cosmological history that includes an early-time component of EDE and a late-time dynamical DE phase.~We show that the full trajectory must pass through several distinct dynamical regimes, namely an initial frozen phase, a transient EDE episode, a subsequent depletion, a prolonged quiescent phase, and a late-time reactivation.~These phases are not independent:~the timing of the early-time release, the amplitude and duration of the EDE phase, the efficiency of the depletion, and the conditions required for late-time reactivation must all arise from the same underlying microphysics.~Consequently, each stage imposes a corresponding set of semi-quantitative, model-independent constraints on the energy density and equation of state of the scalar.

Second, we adopt a complementary top-down approach, embedding these features in scalar field models admitting tracking-like dynamics.~These provide a natural setting for a unified cosmological history, as distinct phases correspond to fixed points of a dynamical system, with the evolution largely insensitive to initial conditions.~In this framework, the model-independent dynamical conditions identified above translate into constraints on the phase-space structure.~We find that these dynamical conditions can only be realized if the potential exhibits three distinct slopes:~an initial slope associated with EDE, a steeper slope governing depletion and quiescence, and a shallow slope responsible for late-time acceleration.~Potentials with fewer slopes fail to reproduce the required dynamics:~two slopes cannot dissipate the EDE sufficiently rapidly, while a single slope cannot simultaneously account for both early- and late-time behavior.~We discuss the implications for the underlying microphysics and for the scalar potentials capable of realizing such a structure.

The paper is organized as follows.~Section~\ref{sec.2} introduces the theoretical framework and the cosmological equations governing the scalar-field dynamics at the background level.~Section~\ref{sec.3} derives the model-independent dynamical requirements that a unified early- and late-time DE history must satisfy.~Section~\ref{sec.4} analyzes minimal single field realizations admitting tracking-like dynamics and identifies the corresponding obstructions to realizing these conditions in phase space.~Section~\ref{sec.5} determines the minimal phase-space structure required to overcome these obstructions and discusses its implications for model building.~Section~\ref{sec.6} summarizes our results and presents our conclusions.~

\section{Theoretical framework}
\label{sec.2}

We consider the action:
\begin{equation}
S = \int d^4x \sqrt{-g} \left[ \frac{M_{\rm Pl}^2}{2} R
- \frac{1}{2} \nabla_\mu \phi \nabla^\mu \phi
- V(\phi) \right]
+ S_{\rm c}[\psi_{\rm c}, A^2(\phi) g_{\mu\nu}]
+ S_{\rm m}[\psi_{\rm m}, g_{\mu\nu}] \, ,
\end{equation}
where $\phi$ is a scalar field with potential $V(\phi)$, the cold dark matter sector, described by $S_{\rm c}[\psi_{\rm c}, A^2(\phi) g_{\mu\nu}]$, is coupled to $\phi$ through the conformal factor $A(\phi)$, while all remaining species collected in $S_{\rm m}[\psi_{\rm m}, g_{\mu\nu}]$ are minimally coupled to $g_{\mu\nu}$.~Hereafter, we work in units where {the speed of light and} the reduced Planck mass {are} set to unity, $M_{\rm Pl}=1=c$.~We  parametrize the coupling between $\phi$ and the cold dark matter sector by
\begin{equation}
\beta(\phi) \equiv \frac{d \ln A(\phi)}{d\phi} \, ,
\end{equation}
so that the minimally coupled limit is recovered for $\beta(\phi)=0$, corresponding to constant $A(\phi)=A_0$.~We set $A_0=1$ without loss of generality.~When the coupling is active, it induces an exchange of energy and momentum within the dark sector, thereby modifying both the background and perturbation evolution of the scalar field and the cold dark matter component.

At the background level, assuming a spatially flat FLRW metric, the cosmological evolution is governed by the Friedmann equations
\begin{equation}
3H^2 = \rho_r + \rho_b + \rho_c + \rho_\phi \, ,
\end{equation}
\begin{equation}
2\dot H + 3H^2 = -\left(p_r + p_b + p_c + p_\phi\right) \, .
\end{equation}
Here $\rho_r$ ($p_r$), $\rho_b$ ($p_b$), and $\rho_c$ ($p_c$) denote the energy densities (pressures) of radiation, baryons, and cold dark matter, respectively, while $\rho_\phi$ and $p_\phi$ are the energy density and pressure of the scalar, given by
\begin{equation}
\rho_\phi = \frac{1}{2}\dot\phi^2 + V(\phi) \qquad\textrm{and}\ 
\qquad
p_\phi = \frac{1}{2}\dot\phi^2 - V(\phi)
\end{equation}
respectively.~The evolution of the different components is then governed by the corresponding continuity equations.~Radiation and baryons satisfy the standard uncoupled continuity equations,
\begin{equation}
\dot\rho_r + 4H\rho_r = 0 \, ,
\qquad
\dot\rho_b + 3H\rho_b = 0 \, ,
\end{equation}
while the coupled dark matter and scalar-field sectors obey
\begin{equation}
\dot\rho_c + 3H\rho_c = - \beta(\phi)\,\dot\phi\,\rho_c \, ,
\end{equation}
\begin{equation}
\dot\rho_\phi + 3H(\rho_\phi+p_\phi) = + \beta(\phi)\,\dot\phi\,\rho_c \, .
\end{equation}
Finally, the dynamics of the scalar field is governed by the Klein-Gordon (KG) equation,
\begin{equation}
\ddot\phi + 3H\dot\phi + V_{,\phi} = \beta(\phi)\,\rho_c \, .
\label{eq:KG}
\end{equation}
In what follows, it is convenient to express the scalar evolution in terms of an \textit{effective potential},
\begin{equation}
V^{\rm eff}(\phi) = V(\phi) + \rho_c \ln A(\phi),
\end{equation}
such that the Klein–Gordon equation takes the form
\begin{equation}
\ddot{\phi} + 3H\dot{\phi} + V^{\rm eff}_{,\phi} = 0.
\end{equation}
We also define the corresponding effective mass:
\begin{equation}
m_{\rm eff}^2 \equiv V^{\rm eff}_{,\phi\phi}.
\end{equation}
In the analysis below we focus on the constant-coupling case, $\beta={\rm const}$, which corresponds to the standard exponential conformal coupling $A(\phi)\propto e^{\beta\phi}$. This choice is commonly used in autonomous-system studies of coupled quintessence because it keeps the system closed and allows the strength of the coupling to be characterized by a single parameter \cite{Copeland:2006wr,Bahamonde:2017ize}. More general choices of $A(\phi)$, such as $A(\phi)\propto\phi$, lead to field-dependent effective couplings $\beta(\phi)$ that lie outside the constant-coupling framework considered here.

\section{Dynamical requirements for a unified early- and late-time dark energy}
\label{sec.3}

\begin{figure}[htpb!]
    \centering
    \includegraphics[width=\linewidth]{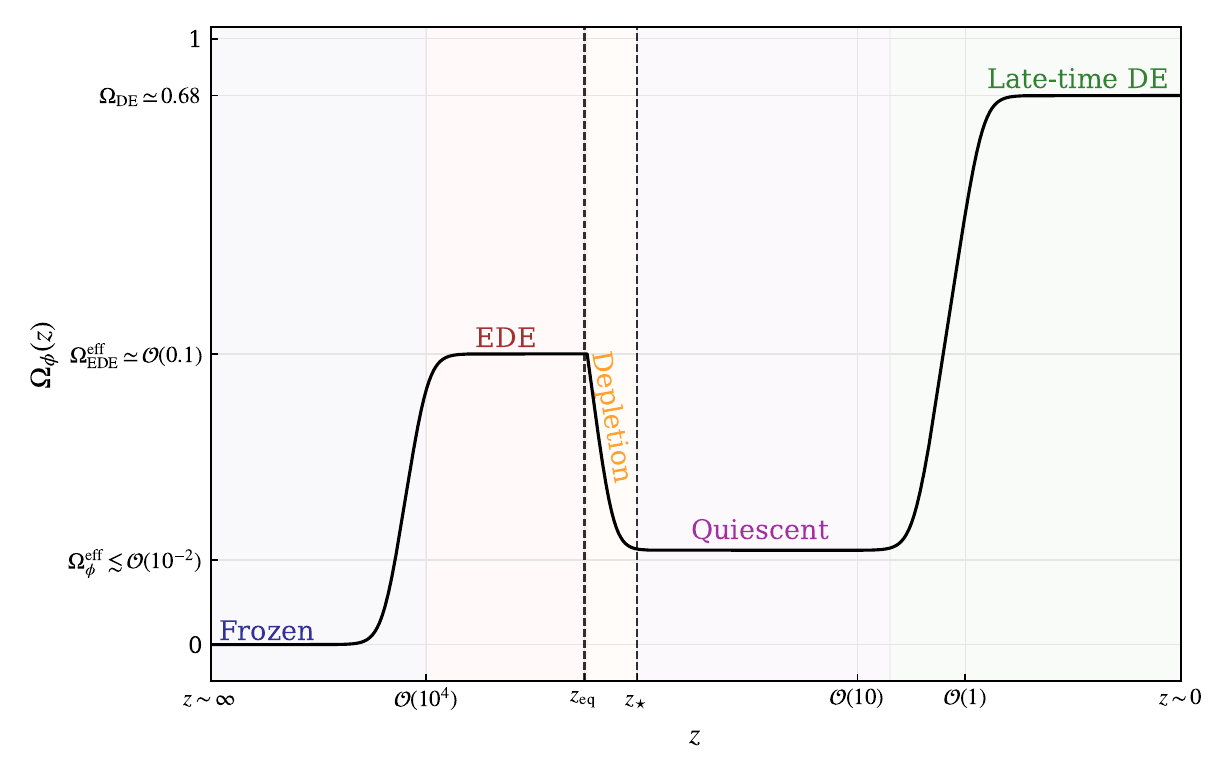}
    \caption{Qualitative evolution of the scalar-field fractional energy density $\Omega_\phi(z)$ in a unified early- and late-time DE scenario.~The figure illustrates the sequence of dynamical regimes that the field must undergo in order to satisfy the requirements discussed in this section:~an initial frozen phase at very high redshift, an EDE phase active around and before matter-radiation equality ($z_{\rm eq}$), a subsequent rapid depletion of the scalar energy density before recombination ($z_{\star}$), a prolonged quiescent phase throughout most of the post-recombination epoch, and a final late-time reactivation responsible for the present accelerated expansion.}
    \label{fig:target_EDE}
\end{figure}

In this section we identify necessary dynamical requirements that a scalar must satisfy in order to account for both early and late DE, while remaining compatible with current observational constraints. Our approach is to characterize the sequence of dynamical regimes that the scalar field must undergo and translate each of them into semi-quantitative constraints on its evolution, without focusing on specific realizations. The resulting conditions are therefore necessary, but not sufficient: while they constrain the background trajectory required for a unified early- and late-time DE model, they do not capture the additional constraints that arise when perturbations are included. Models satisfying these conditions must therefore still be confronted with cosmological data through dedicated parameter-inference analyses to determine their viability.

To this end, we describe the scalar-field dynamics in terms of the fractional energy density $\Omega_\phi$ and the corresponding equation of state $w_\phi$:
\begin{equation}
\Omega_\phi \equiv \frac{\rho_\phi}{\rho_{\rm tot}} \quad, \quad
w_\phi \equiv \frac{p_\phi}{\rho_\phi}.
\end{equation}

We show that a successful realization requires the field to pass through five distinct phases:~an initial frozen regime, a transient EDE phase around matter-radiation equality, a rapid depletion, a prolonged quiescent period after recombination, and finally a late-time reactivation responsible for the present-day accelerated expansion.~This sequence is illustrated schematically in Fig.~\ref{fig:target_EDE}.~We now analyze each of these regimes in turn.

\subsection{Frozen phase}

In the early Universe, deep in the radiation  era, the scalar must remain subdominant to preserve the standard expansion history and thermal evolution i.e., $\Omega_\phi \ll \Omega_r \simeq 1$.~This is naturally accomplished if the field is overdamped with $\dot{\phi} \simeq 0$, implying the equation of state $w_\phi \simeq -1$.~This behavior arises whenever the effective mass of the scalar is small compared with the Hubble scale, $m_{\rm eff}^2 \ll H^2$.~The frozen phase thus provides natural initial conditions since the Hubble parameter increases in the past.

\subsection{Early Dark Energy phase}
\label{sec.EDE}

To address the Hubble tension, the frozen phase must end before recombination, allowing the scalar to temporarily contribute as EDE.~Below, we briefly review the role of EDE in resolving the Hubble tension before deriving the dynamical conditions required for it to operate efficiently.~We then translate these conditions into constraints on a scalar field capable of realizing an EDE phase.

\subsubsection{Early dark energy}

EDE alleviates the Hubble tension by reducing the sound horizon at recombination, defined as
\begin{equation}
r_s(z_\star)=\int_{z_\star}^{\infty}\frac{c_s(z)}{H(z)}\,dz\qquad\textrm{with}\qquad H^2(z)=\frac{1}{3} \left[ \rho_r(z)+\rho_m(z)+\rho_{\rm EDE}(z)\right]\,\,, 
\label{eq:rs}
\end{equation}
where we have neglected late DE since it is negligible.~The EDE component becomes dynamically relevant near matter--radiation equality (see Fig.~\ref{fig:target_EDE}), increasing $H(z)$ relative to its $\Lambda$CDM value over the relevant pre-recombination interval and thereby reducing $r_s(z_\star)$.~Ultimately, the inferred value of $H_0$ is increased to preserve the geometric relation between $r_s(z_{\star})$, the angular diameter distance to the CMB, $D_A(z_{\star})$, and the angular scale of the acoustic peaks $\theta_\star$, given by
\begin{equation}
\theta_\star = \frac{r_s(z_\star)}{D_A(z_\star)}.
\end{equation}
The angle $\theta_\star$ is constrained by CMB observations to a few parts in $10^4$~\cite{SPT-3G:2025bzu}, making it a powerful geometric anchor that must be preserved with high accuracy by any viable modification of the pre- and/or post-recombination expansion history.~Consequently, any change in the sound horizon must be compensated by a corresponding change in the angular diameter distance, given by 
\begin{equation}
\label{eq:DA}
D_A(z_\star)=\frac{1}{1+z_\star}
\int_0^{z_\star}
\frac{dz}{H(z)}=\frac{H_0^{-1}}{1+z_\star}
\int_0^{z_\star}
\frac{dz}{E(z)},
\end{equation}
{where $E(z)=H(z)/H_0$.} 
~To leading order one has $\delta r_s/r_s \simeq \delta D_A / D_A$.~Since, at fixed cosmological parameters, $D_A(z_\star) \propto H_0^{-1}$, this implies $\delta r_s/r_s \simeq -\delta H_0/H_0$.~Shifting the Hubble constant from the CMB-inferred value  $H_0 \sim 67$ km/s/Mpc to the locally measured 
$H_0 \sim 73$ km/s/Mpc corresponds to a relative shift of order $\delta H_0 / H_0 \sim 0.08$ which translates into a required reduction of the sound horizon at the level of approximately $8\%$.~We refer the reader to Refs.~\cite{Kamionkowski:2022pkx,Poulin:2025nfb} for more details of EDE.

\subsubsection{Conditions for Early Dark Energy to resolve the Hubble tension}

\begin{figure}[t]
    \centering
    \includegraphics[width=0.7\linewidth]{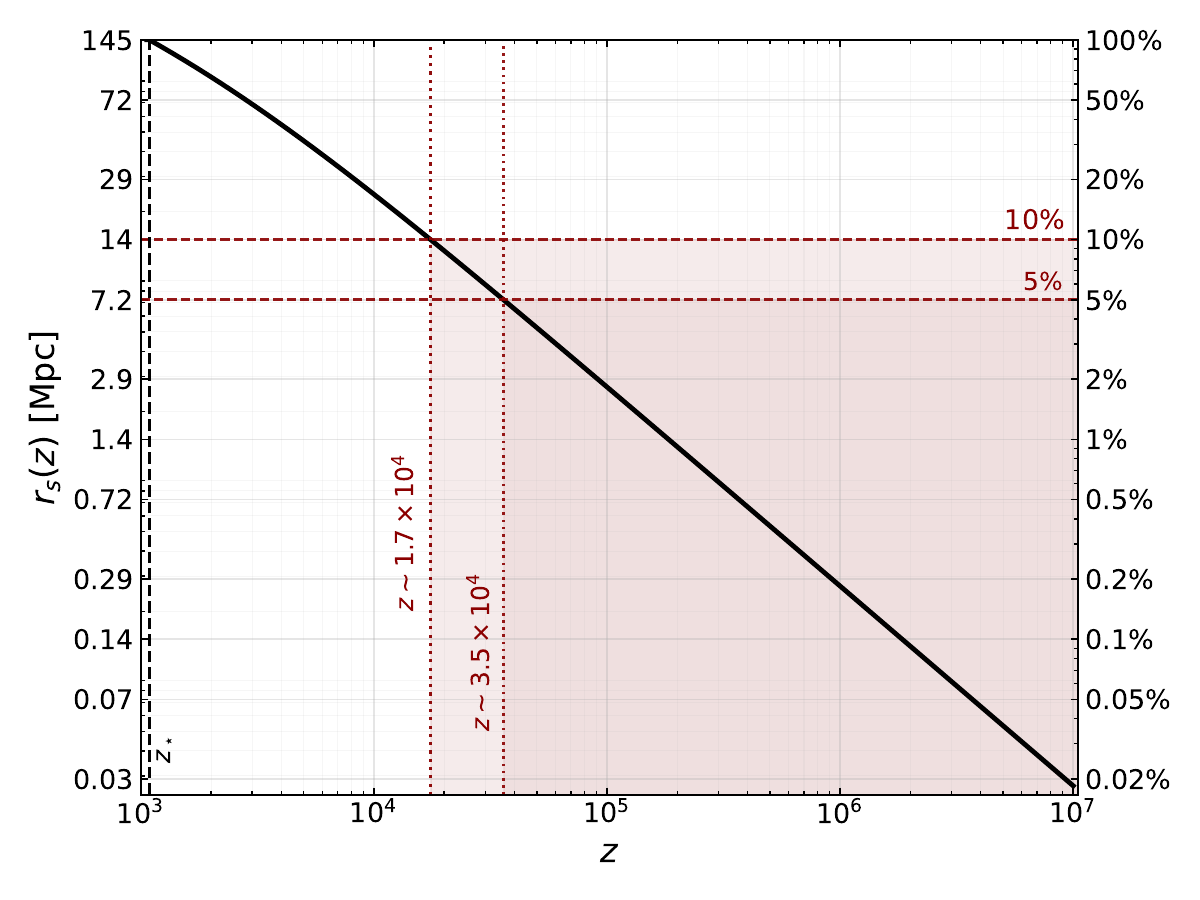}
    \caption{Evolution of $r_s(z)$ for a Planck best-fit $\Lambda$CDM cosmology.~The left axis shows the value of $r_s(z)$ in Mpc, while the right axis shows the ratio $r_s(z)/r_s(z_\star)$.~The vertical dashed line marks recombination at $z_\star \simeq 1090$.~The horizontal dashed lines indicate the $10\%$ and $5\%$ levels of $r_s(z_\star)$, and the corresponding vertical dotted lines mark the redshifts at which $r_s(z)$ reaches those values.}
    \label{fig:rs}
\end{figure}

In order to resolve the Hubble tension, EDE must become relevant at redshifts where it can efficiently modify the sound horizon $r_s(z_\star)$ at the $\sim 5$--$10\%$ level via Eq.~\eqref{eq:rs}.~To quantify this, we consider the contribution to the sound horizon accumulated from a given redshift $z$ to arbitrarily early times.~In Fig.~\ref{fig:rs} we show the residual contribution to the total sound horizon $r_s(z_\star)$,
\begin{equation}
r_s(z)=\int_z^{\infty}\frac{c_s(z')}{H(z')}\,dz',
\label{eq:rs(z)}
\end{equation}
evaluated in $\Lambda$CDM.~As indicated in the figure, only $\approx 10\%$ of the total sound horizon is sourced by redshifts $z \gtrsim 1.7\times 10^4$, and only about $5\%$ by $z \gtrsim 3.5\times 10^4$.~Modifications to the expansion history at earlier times are therefore increasingly inefficient { and cannot fully resolve the tension on their own if localized only at $z\gtrsim 10^4$.} In addition, $\Omega\ede$ must become subdominant by recombination in order to preserve {consistency with CMB data.}~Taken together, these considerations suggest that viable EDE scenarios must be relevant at redshifts $z_\star \lesssim z \lesssim \mathrm{few}\times 10^4$.

We can estimate the amount of EDE needed to reduce $r_s$ sufficiently by considering it as a perturbation about $\Lambda$CDM.~Let $H_{\rm bg}(z)$ be the Hubble rate in the absence of EDE.~The full Hubble rate is  $H^2(z)= H_{\rm bg}^2(z) \left[1-\Omega\ede(z)\right]^{-1}$.~Expanding to first order in $\Omega\ede$, one finds $\delta H /H \simeq 1 /2\,\Omega\ede(z)$.~The corresponding fractional variation of the sound horizon can then be written schematically as
\begin{equation}
\frac{\delta r_s}{r_s}
\simeq
-\int_{z_\star}^{\infty} d\ln z\, W_{r_s}(z)\,\frac{\delta H(z)}{H(z)}
\simeq
-\frac{1}{2}\int_{z_\star}^{\infty} d\ln z\, W_{r_s}(z)\,\Omega\ede(z) \, ,
\end{equation}
where $W_{r_s}(z)\equiv z\,c_s(z)/\left[H(z)r_s(z_\star)\right]$ is a normalized weight function describing the fractional contribution of each logarithmic redshift interval to the sound-horizon integral.~As discussed above, EDE primarily contributes to the sound horizon reduction at redshifts $z\lesssim z_{\rm cut}\sim {\rm few}\times 10^4$, so we can define the effective weighted average
\begin{equation}
\Omega\ede^{\rm eff}\equiv
\int_{z_\star}^{z_{\rm cut}} d\ln z\, W_{r_s}(z)\,\Omega\ede(z)
\end{equation}
One then has $\delta r_s / r_s \simeq -1/2\, \Omega_\phi^{\rm eff}$.~Therefore, a reduction of the sound horizon at the level required to alleviate the Hubble tension, $\delta r_s / r_s \simeq 5-10\%$ corresponds to an effective EDE contribution of order $\Omega\ede^{\rm eff}\sim 0.1$--$0.2$.

\subsubsection{Conditions for a scalar to realize an Early Dark Energy phase}

The results above imply that scalar models of EDE must become relevant after $z\simeq{\rm few}\times 10^4$ and contribute an effective $\Omega^{\rm eff}_\phi=\Omega\ede^{\rm eff}\simeq0.1$ before~{rapidly redshifting away} by recombination at $z=z_\star\simeq1100$.~This implies the field must exit the frozen regime and begin rolling during this period, so its effective mass must satisfy $H(z_\star)\lesssim m_{\rm eff}\lesssim H(z\simeq10^4)$, which imposes 
\begin{equation}
m_{\rm eff}\sim 10^{-28} - 10^{-26}\ {\rm eV}.
\end{equation}

\subsection{Depletion phase}

As noted above, the EDE phase must end by recombination because the CMB is extremely sensitive to the state of the Universe at decoupling.~Modifications of the background expansion rate at $z_\star$ shift the acoustic scale, while residual scalar perturbations leave direct imprints on the temperature and polarization anisotropy spectra.~Both effects are tightly constrained by CMB observations at the percent level~\cite{SPT-3G:2025vyw}.~This implies that the scalar must lose energy sufficiently rapidly to avoid violating these constraints.

To quantify this requirement, it is convenient to consider the ratio $\Omega_\phi(z_\star)/\Omega_\phi^{\rm eff}$, which measures the amount by which the scalar contribution is diluted after the EDE phase. ~In order to preserve agreement with CMB data, one typically requires 
\begin{equation}
\Omega_\phi(z_\star)/\Omega_\phi^{\rm eff} \lesssim 10^{-1} \, .
\end{equation}
This estimate follows from two considerations.~First, CMB observations  constrain deviations from the $\Lambda$CDM expansion history at recombination at the percent level, implying $\Omega_\phi(z_\star) \lesssim \mathcal{O}(0.01)$~\cite{SPT-3G:2025bzu,SPT-3G:2025vyw}.~Second, as derived above, alleviating the Hubble tension through an early-time mechanism requires a fractional energy density  $\Omega_\phi^{\rm eff} \sim \mathcal{O}(0.1)$.~

\begin{figure}[t!]
    \centering
    \includegraphics[width=0.9\linewidth]{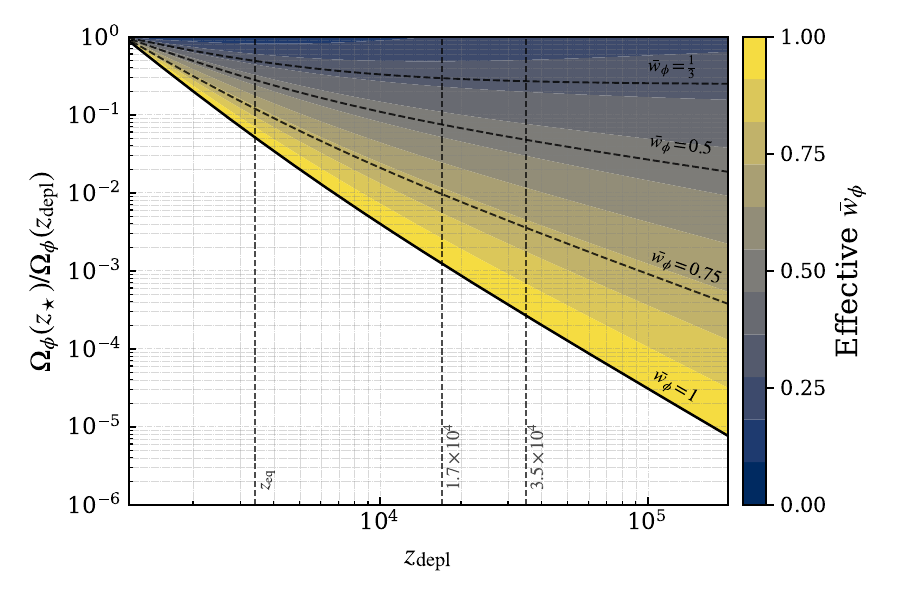}
    \caption{Depletion ratio $\Omega_\phi(z_\star)/\Omega_\phi(z_{\rm depl})$ as a function of the depletion redshift $z_{\rm depl}$, which marks the onset of the depletion phase.~The color scale encodes the effective equation of state $\bar{w}_\phi$ governing the evolution of the scalar field, while the dashed contours highlight representative values.~The vertical dashed lines indicate, from left to right, matter-radiation equality ($z_{\rm eq}$) and the redshift inferred in Section~\ref{sec.EDE}, above which the reduction of the sound horizon becomes strongly geometrically inefficient.~For clarity, $\bar{w}_\phi$ is restricted to the range $0 \le \bar{w}_\phi \le 1$, corresponding to canonical scalar-field dynamics.}
    \label{fig:depletion}
\end{figure}

The depletion phase is therefore characterized by a rapid reduction of the scalar energy density between some redshift $z_{\rm depl}$ and $z_\star$. Since the EDE maximum cannot occur arbitrarily far from recombination, this interval is relatively limited, and the post-peak evolution can be conveniently described by an effective equation of state $\bar{w}_\phi$, representing the average of $w_\phi$ over this range.~Assuming $\bar{w}_\phi$ is approximately constant, the scalar energy density scales as $\rho_\phi(z)\propto(1+z)^{3(1+\bar{w}_\phi)}$, while the total background remains well approximated by $\rho_{\rm tot}(z)\simeq \rho_r(z)+\rho_m(z)$, consistently with $\Omega_\phi$ rapidly becoming subdominant after the peak.~It follows that
\begin{equation}
\frac{\Omega_\phi(z_\star)}{\Omega_\phi(z_{\rm depl})}
\simeq 
\left(\frac{1+z_\star}{1+z_{\rm depl}}\right)^{3(1+\bar{w}_\phi)}
\frac{\rho_{\rm tot}(z_{\rm depl})}{\rho_{\rm tot}(z_\star)} \, .
\label{eq:depletion_ratio}
\end{equation}

In Fig.~\ref{fig:depletion} we show the depletion ratio $\Omega_\phi(z_\star)/\Omega_\phi(z_{\rm depl})$ as a function of the depletion redshift $z_{\rm depl}$, with the effective equation of state $\bar{w}_\phi$ encoded by the color scale.~For each pair $(z_{\rm depl},\,\Omega_\phi(z_\star)/\Omega_\phi(z_{\rm depl}))$, the color identifies the value of $\bar{w}_\phi$ required to achieve that level of suppression, while the dashed lines highlight representative values.~Moving along the horizontal axis corresponds to shifting the location of the EDE maximum in redshift.~As $z_{\rm depl}$ is pushed to higher redshift, the scalar field has a longer interval over which to dilute before decoupling, and the required value of $\bar{w}_\phi$ decreases accordingly.~However, this trend is bounded by the timing constraint derived in the previous section.~The vertical lines mark the redshifts beyond which modifications to the expansion history become inefficient, contributing less than $5$--$10\%$ to $r_s(z_\star)$.~The viable region is therefore restricted to $z_{\rm depl} \lesssim \mathrm{few}\times 10^4$, and in practice lies closer to matter-radiation equality, where the depletion must occur over a limited redshift interval.~In this regime, even a relatively efficient dilution leads only to a moderate suppression of the scalar contribution at $z_\star$.~{If depletion begins around matter-radiation equality, achieving the necessary suppression generally requires $\bar{w}_\phi \gtrsim 0.5$.} \footnote{{This is consistent with axion-like EDE where, during oscillations around the minimum, the averaged equation of state is $\bar{w}_\phi=(n-1)/(n+1)$, with $n$ controlling the power-law behavior of the potential near the minimum. To ensure sufficiently rapid depletion, a common choice is $n=3$, which gives $\bar{w}_\phi=1/2$~\cite{Poulin:2025nfb}.}}

This discussion raises a crucial distinction between canonical EDE and unified models of early and late DE.~In the former, the depletion phase is driven by oscillations around a minimum of the potential, which lead to a rapid redshifting of the scalar energy density.~While this efficiently suppresses the scalar contribution, it also drives the field toward the minimum, where it subsequently remains so that it cannot act as dynamical late DE.~Thus, the depletion phase represents a novel challenge for unified scenarios:~the scalar must dilute sufficiently rapidly to become negligible by recombination while retaining the possibility of late-time reactivation.

\subsection{Quiescent phase}
A second challenge for unified early- and late-time DE scenarios is that, post-depletion, the field must experience a prolonged quiescent phase to ensure that the reduced sound horizon is compensated by an increase in $H_0$ and not new late-time physics.~Specifically, since $\theta_\star=r_s(z_\star)/D_A(z_\star)$, a scalar active post-recombination could compensate for the reduced $r_s$ by altering the integrand in Eq.~\eqref{eq:DA} rather than by increasing $H_0$.

\begin{figure}[tpb!]
    \centering
    \includegraphics[width=0.8\linewidth]{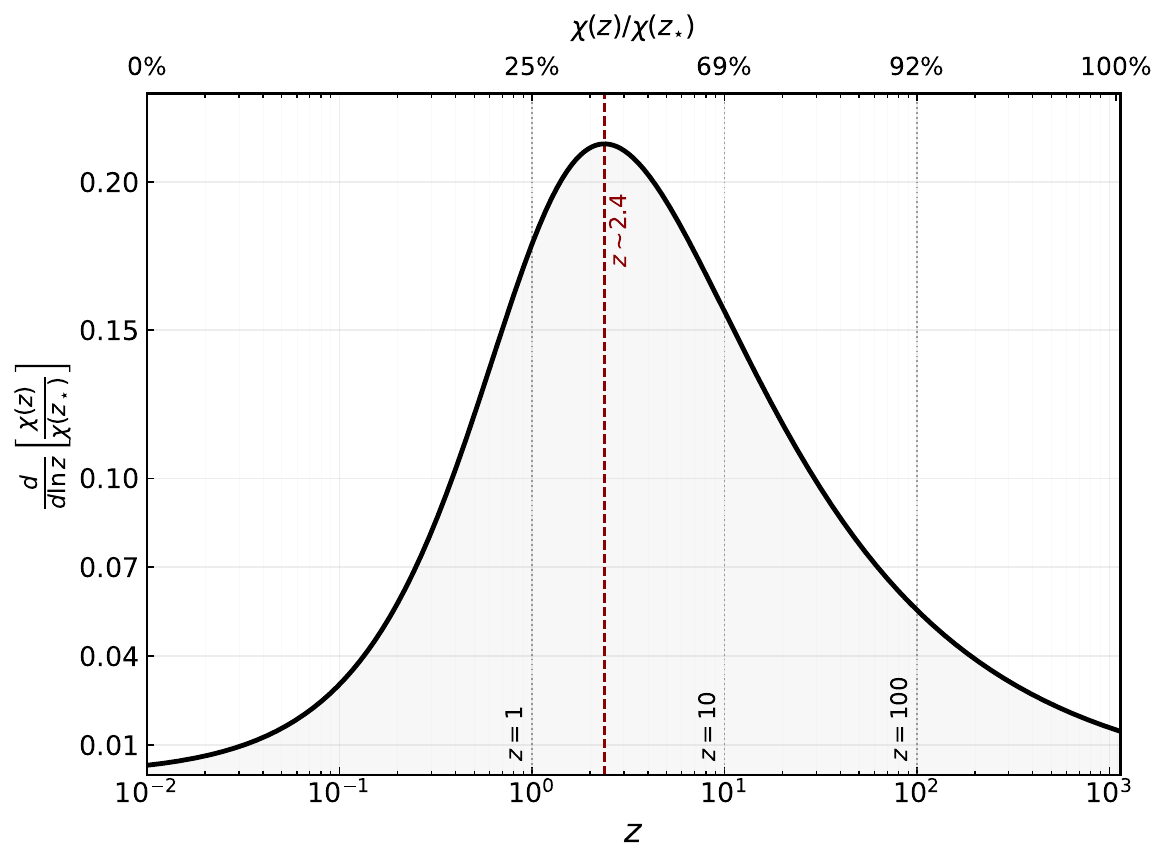}
    \caption{Differential contribution of each logarithmic redshift interval to the normalized comoving distance $\chi(z_\star)$ for a Planck best-fit $\Lambda$CDM cosmology.~The black curve shows $d[\chi(z)/\chi(z_\star)]/d\ln z$ as a function of redshift, while the upper axis reports the corresponding cumulative fraction $\chi(z)/\chi(z_\star)$.~The differential contribution peaks broadly around $z\simeq 2.4$, marked by the red dashed vertical line, and remains sizable over a wide post-recombination interval.~}
    \label{fig:chi}
\end{figure}

To quantify this requirement, it is useful to work in terms of the comoving distance
\begin{equation}
\chi(z_\star)=\int_0^{z_\star}\frac{dz}{H(z)} \, ,
\end{equation}
which is directly related to the angular diameter distance via $D_A(z_\star)=\chi(z_\star)/(1+z_\star)$.~This quantity encodes the full post-recombination expansion history, and any residual scalar contribution affects it through its impact on $H(z)$ over an extended redshift range.~Its sensitivity to different redshift intervals is captured by the differential contribution to the normalized comoving distance per logarithmic redshift interval,
\begin{equation}
\frac{d}{d\ln z}\left[\frac{\chi(z)}{\chi(z_\star)}\right] \, ,
\end{equation}
shown in Fig.~\ref{fig:chi}, together with the cumulative fraction $\chi(z)/\chi(z_\star)$ on the upper axis.\footnote{We formulate the differential relative contribution in terms of $\chi(z)/\chi(z_\star)$ because the running ratio $D_A(z)/D_A(z_\star)$ would introduce an additional factor $(1+z_\star)/(1+z)$, which is not part of the line-of-sight integral itself and would artificially suppress the contribution from higher redshifts.
} As seen from the figure, the differential weight peaks broadly around $z\sim 2.4$ and remains sizable over a wide redshift interval.~This behavior is qualitatively different from that of the sound-horizon integral, which is dominated by a more localized region near recombination, implying that the post-recombination geometry is sensitive to a much broader range of redshifts.~Consequently, even a relatively modest scalar contribution after recombination can leave a non-negligible imprint on $\chi(z_\star)$.~

\begin{figure}[tpb!]
    \centering
    \includegraphics[width=0.9\linewidth]{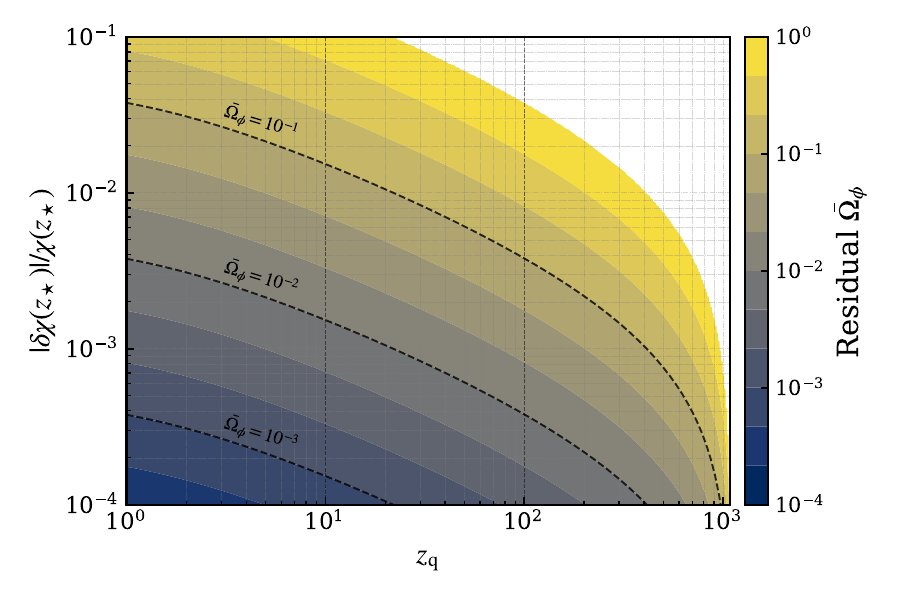}
    \caption{Relative correction $|\delta\chi(z_\star)/\chi(z_\star)|$ of the comoving distance induced by a nearly constant residual contribution $\bar{\Omega}_\phi$ persisting from recombination, $z\simeq z_\star$, down to a redshift $z\simeq z_{\rm q}$.~The horizontal axis shows different values of $z_{\rm q}\in[z_{\star}\, , \, 1]$, while the color scale encodes the value of $\bar{\Omega}_\phi$.~The figure illustrates how the geometric constraint imposed by the quiescent phase depends both on the amplitude of the residual post-recombination fraction and on the redshift interval over which it persists.}
    \label{fig:quiescence}
\end{figure}

As established in the previous section, the scalar field must be subdominant by decoupling so we can perturb in $\Omega_\phi$ so that $\delta H/H \simeq \frac{1}{2}\Omega_\phi(z)$, which gives
\begin{equation}
\frac{\delta\chi(z_\star)}{\chi(z_\star)}
\simeq
-\frac{1}{2}\int_0^{z_\star} d\ln z\, W_\chi(z)\,\Omega_\phi(z)\equiv-\frac12\langle \Omega_\phi\rangle,
\label{eq:quiescence}
\end{equation}
where $W_\chi(z)$ is the normalized geometric kernel shown in Fig.~\ref{fig:chi}.~Since  EDE reduces $r_s(z_\star)$ by $\sim 5-10\%$, we require $\delta\chi(z_\star)/\chi(z_\star)$ to remain at the sub-percent level.~This constraint  applies to the weighted average of $\Omega_\phi(z)$ over the entire post-recombination history, $\langle\Omega_\phi\rangle$.~

Of particular relevance to the models we will study is the case where, after recombination, the field constitutes an approximately constant fraction $\bar{\Omega}_\phi$ over a redshift interval extending from $z_\star$  to some $z_{\rm{q}}\ll z_\star$.~In this scenario, Eq.~\eqref{eq:quiescence} reduces to
\begin{equation}
\left|\frac{\delta\chi(z_\star)}{\chi(z_\star)}\right|
\simeq
\frac{1}{2}\,\bar{\Omega}_\phi
\int_{z_{\rm{q}}}^{z_\star} d\ln z\, W_\chi(z)\,,
\label{eq:quiescence_plateau}
\end{equation}
revealing that the geometric impact of the residual scalar contribution is controlled by two factors:~the level at which the post-recombination plateau is maintained, and the redshift interval over which it persists.~This is illustrated in Fig.~\ref{fig:quiescence}, where we show $|\delta\chi(z_\star)/\chi(z_\star)|$ as a function of $z_{\rm{q}}$ for various values of $\bar{\Omega}_\phi$.~The figure provides a useful quantitative characterization of the quiescent phase:~maintaining $\bar{\Omega}_\phi \lesssim \mathcal{O}(10^{-2})$ after recombination is sufficient to keep the induced correction to $\chi(z_\star)$ safely subdominant.~In contrast, residual plateaus at the level of several percent or more reduce $D_A(z_\star)$ directly, so that the reduction in $r_s(z_\star)$ is compensated by the post-recombination scalar contribution rather than by a larger inferred value of $H_0$.

The quiescence requirement is naturally compatible with the depletion condition derived above.~If the scalar contribution has already been reduced to the percent level by recombination, then maintaining it at or below that level throughout the subsequent evolution is sufficient to preserve the post-recombination distance geometry.~Thus, efficient depletion opens the window for a viable quiescent phase, provided the scalar does not reactivate too early.

The quiescent phase therefore imposes an extended constraint on the scalar dynamics.~Unlike the EDE phase, which requires a localized increase of $H(z)$ over a relatively narrow redshift window, quiescence requires the absence of any non-negligible modification to the expansion rate throughout a much broader interval, spanning a substantial fraction of the post-recombination history.

\subsection{Late-time Dark Energy phase}

After remaining subdominant throughout the post-recombination epoch, the same degree of freedom responsible for the early-time dynamics must eventually account for the late-time DE component.~At the background level, this requires:
\begin{equation}
\Omega_\phi(z=0)\sim 0.68\,,\qquad w_\phi(z\lesssim 1)< -1/3,
\end{equation}
which introduces a further dynamical requirement:~the scalar must reactivate only at late times and grow from a negligible contribution to $\Omega_\phi(z=0)\sim 0.68$.~Since the contribution to $\chi(z_\star)$ remains largest around $z\sim 2.4$ and sizable over a broad redshift interval, premature reactivation would violate the quiescence requirement.~The onset of late-time growth must therefore occur at $z\lesssim \mathcal{O}(\mathrm{few})$, forcing the transition to take place over a relatively limited interval, $\Delta z\lesssim \mathcal{O}(1)$--$\mathcal{O}(\mathrm{few})$.~This requirement is especially restrictive because the evolution is constrained not only by integrated quantities such as $\chi(z_\star)$, but also by direct measurements of $H(z)$ and distances from BAO and Type Ia supernovae across $z\lesssim 3$~\cite{eBOSS:2020yzd,Scolnic:2021amr,Brout:2022vxf,DESI:2025zgx,DESI:2024uvr,DESI:2024lzq,DES:2025sig,Rubin:2026qdt}.~It also poses a coincidence problem:~after remaining subdominant over most of cosmic history, the field must become relevant precisely near the epoch of DE domination.~Late-time attractor or tracking behavior can naturally address this timing issue; without it, the onset of acceleration would depend sensitively on the state inherited from the EDE, depletion, and quiescent phases.

The late-time DE phase thus closes the full dynamical sequence imposed on the scalar field.~In the following section, we assess whether minimal single-field realizations can satisfy all of these requirements simultaneously, and identify the obstructions that arise when they cannot.

\section{Minimal single-field realizations}
\label{sec.4}

In Section~\ref{sec.3} we derived a series of stringent dynamical requirements that a single scalar must satisfy to account for both early- and late-time DE.~Specifically, the field must remain subdominant in the early universe, undergo a transient phase of non-negligible energy density around matter-radiation equality, rapidly lose energy before recombination, remain quiescent for an extended period, and eventually reproduce the observed late-time accelerated expansion.~See Fig.~\ref{fig:target_EDE}.~In this section, we translate these requirements into properties of the potential and coupling for the class of minimally and conformally coupled quintessence models introduced in Section~\ref{sec.2}.~While our analysis remains at the background level, perturbations are crucial in EDE model building. For example, monomial thawing models such as the $\phi^6$ Rock 'n' Roll EDE model can produce an EDE-like background evolution, but are less successful once their full perturbation evolution is confronted with cosmological data~\cite{Agrawal:2019lmo}.

To remain as model-independent as possible, we formulate the problem in the language of autonomous dynamical systems \cite{Copeland:1997et,Copeland:2006wr,Bahamonde:2017ize}.~This framework captures the phase-space structure of a broad class of potentials and allows us to identify the fixed points associated with the dynamical requirements derived above. ~General potentials do not, in general, lead to closed autonomous systems, but the fixed-point structure still provides a useful local description whenever the logarithmic slope
\begin{equation}
\lambda \equiv -\frac{V_{,\phi}}{V}\,,
\label{eq:lambdadef}
\end{equation}  
varies slowly over the relevant portion of the scalar field evolution.~This formalism is therefore most appropriate for potentials whose evolution can be approximated by successive quasi-autonomous regimes with nearly constant $\lambda$, as in exponential, tracking, or quasi-exponential regions. Potentials involving rapid, non-adiabatic variations of $\lambda$ may not reach the fixed points and require a separate analysis.~Since the qualitative fixed-point structure is typically insensitive to precise parameter choices, a model whose phase space does not contain the required sequence of dynamical regimes cannot reproduce the desired phenomenology.~The dynamical-systems analysis therefore provides a diagnostic of whether a given potential can support unified early- and late-time DE behavior and, when it cannot, reveals the structural obstruction that prevents it.

We begin by introducing the general phase-space formalism.~We then apply this framework to increasingly complex realizations, assessing whether they can reproduce the sequence of dynamical regimes required by a unified early- and late-time DE scenario.~This allows us to identify the obstructions that arise in minimal models and to determine the structure required to overcome them.

\subsection{Phase-space formalism}

To cast the cosmological evolution into autonomous form, we introduce the dimensionless variables
\begin{equation}
x \equiv \frac{\dot{\phi}}{\sqrt{6}\,H}\,,
\qquad
y \equiv \frac{\sqrt{V}}{\sqrt{3}\,H}\,,
\qquad
\Omega_r \equiv \frac{\rho_r}{3H^2}\,,
\qquad
\Omega_b \equiv \frac{\rho_b}{3H^2}\,.
\label{eq:xydefs}
\end{equation}

The Friedmann constraint then implies
\begin{equation}
\Omega_c = 1 - \Omega_r - \Omega_b - x^2 - y^2\,,
\label{eq:omegac_constraint}
\end{equation}
where $\Omega_c$ denotes the dark matter density fraction.~These variables have a direct physical interpretation.~The variable $x$ measures the relative importance of the field velocity with respect to the Hubble scale and therefore quantifies how kinetic-dominated the scalar dynamics is.~The variable $y$ instead tracks the potential contribution and becomes large when the field behaves more vacuum-like.~Their combination determines the scalar energy density and equation of state:
\begin{equation}
\Omega_\phi = x^2 + y^2\,, 
\qquad
w_\phi = \frac{x^2-y^2}{x^2+y^2}.
\label{eq:wphi_xy}
\end{equation}
The latter interpolates between stiff-like behavior, $w_\phi \simeq 1$, and vacuum-like behavior, $w_\phi \simeq -1$, depending on the relative size of the kinetic and potential terms.

The background dynamics are described by the variables $\left\{x,\,y,\,\Omega_r,\,\Omega_b\right\}$ once a scalar potential is specified.~The dependence on the potential can be encoded  conveniently by the logarithmic slope of the potential, Eq.~\eqref{eq:lambdadef}, which measures the local steepness of the potential along the field trajectory.

Using the e-fold number $N\equiv \ln a$ as time variable, the background equations become
\begin{align}
x' &= -3x + \sqrt{\frac{3}{2}}\,\lambda y^2 + \sqrt{\frac{3}{2}}\,\beta\,\Omega_c
      + \frac{3}{2}x\left(1+x^2-y^2+\frac{\Omega_r}{3}\right)\,,
\label{eq:xprime_generic}
\\
y' &= -\sqrt{\frac{3}{2}}\,\lambda x y
      + \frac{3}{2}y\left(1+x^2-y^2+\frac{\Omega_r}{3}\right)\,,
\label{eq:yprime_generic}
\\
\Omega_r' &= \Omega_r\left(-1+3x^2-3y^2+\Omega_r\right)\,,
\label{eq:omegarprime_generic}
\\
\Omega_b' &= \Omega_b\left(3x^2-3y^2+\Omega_r\right)\,,
\label{eq:omegabprime_generic}
\\
\lambda' &= -\sqrt{6}\,x\,\lambda^2(\Gamma-1)\,.
\label{eq:lambdaprime_generic}
\end{align}
where a prime denotes differentiation with respect to $N$ and 
\begin{equation}
\Gamma \equiv \frac{V\,V_{,\phi\phi}}{V_{,\phi}^2}\,.
\label{eq:Gammadef}
\end{equation}
Equations \eqref{eq:xprime_generic}--\eqref{eq:lambdaprime_generic} form a closed autonomous system provided that $\Gamma$ can be expressed as a function of $\lambda$.~An exponential potential $V(\phi)\sim\exp(-\alpha\phi)$ provides the simplest example, with $\Gamma=1$ and hence constant $\lambda=\alpha$.~More generally, only specific potentials have the property that $\Gamma=\Gamma(\lambda)$.

When the system is autonomous, {its} critical points  are obtained by imposing $x'=y'=\Omega_r'=\Omega_b'=\lambda'=0$.~These correspond to stationary configurations in phase space.~When $\Gamma$ cannot be written as a function of $\lambda$, the system is not closed in the variables above.~If $\lambda$ varies slowly, however, one can treat it as an adiabatic parameter and solve
$x'=y'=\Omega_r'=\Omega_b'=0$
at fixed $\lambda$.~The resulting quasi-static fixed points provide a local description of the evolution, with their properties changing slowly as $\lambda(N)$ evolves.~This approximation is particularly useful for potentials whose slope is approximately constant over extended portions of the trajectory, since the cosmology can then be interpreted as a sequence of locally autonomous systems.

We now investigate how the dynamical requirements derived in Section~\ref{sec.3} can be incorporated into the fixed point structure.~Our strategy is to begin with simple models and add complexity until we identify the minimal set of model features capable of realizing all of the required regimes.

\subsection{Conformally-coupled double exponential}

As a first non-trivial realization, we consider a double-exponential potential\footnote{A single exponential potential contains only one constant-$\lambda$ branch.~Such models cannot provide enough distinct dynamical regimes to support a unified EDE and late-time DE history.}, allowing for a constant coupling $\beta$ to dark matter:
\begin{equation}
V(\phi)=V_1 e^{-\lambda_{\rm early}\phi}+V_2 e^{-\lambda_{\rm late}\phi}\,,
\label{eq:V_double_exp}
\end{equation}
where we take $\lambda_{\rm early}>\lambda_{\rm late}$ without loss of generality.~For this potential, the logarithmic slope becomes
\begin{equation}
\lambda \equiv -\frac{V_{,\phi}}{V}
=
\frac{\lambda_{\rm early}V_1 e^{-\lambda_{\rm early}\phi}+\lambda_{\rm late}V_2 e^{-\lambda_{\rm late}\phi}}
{V_1 e^{-\lambda_{\rm early}\phi}+V_2 e^{-\lambda_{\rm late}\phi}}\,,
\label{eq:lambda_double_exp}
\end{equation}
so that $\lambda$ is not constant, but interpolates dynamically between the two asymptotic slopes $\lambda_{\rm early}$ and $\lambda_{\rm late}$.~The quantity $\Gamma$ defined in Eq.~\eqref{eq:Gammadef} can then be expressed in terms of $\lambda$ as
\begin{equation}
\Gamma-1
=
\frac{(\lambda-\lambda_{\rm early})(\lambda-\lambda_{\rm late})}{\lambda^2}\,.
\label{eq:Gamma_minus_one_double_exp}
\end{equation}
Substituting this result into Eq.~\eqref{eq:lambdaprime_generic}, one finds that critical points with $\lambda'=0$ lie on one of the three branches
\begin{equation}
x=0\,, \qquad \lambda=\lambda_{\rm early}\,, \qquad \lambda=\lambda_{\rm late}\,.
\label{eq:double_exp_branches}
\end{equation}
The branch $x=0$ corresponds to configurations in which the field is frozen, while the other two branches represent regimes in which the field rolls with different logarithmic slopes.~Along these constant-$\lambda$ branches, the system locally behaves as a single-exponential potential with $\lambda=\lambda_{\rm early}$ or $\lambda=\lambda_{\rm late}$, whose critical points are summarized in Table~\ref{tab:fixed_points}.~Given the hierarchy of slopes, and barring finely tuned initial conditions or amplitudes, we expect the $\lambda_{\rm early}$ branch to be realized first, followed by the $\lambda_{\rm late}$ branch.~We therefore identify the former as a candidate for the EDE phase and the latter as a potential late-time DE phase.~By contrast, the branch $x=0$ does not define an independent asymptotic exponential regime, but rather mediates transitions between qualitatively different dynamical behaviors.

\begin{table*}[t]
\centering
\renewcommand{\arraystretch}{2}
\resizebox{0.99\textwidth}{!}{
\begin{tabular}{l | c c c c c | l}
\hline
\textbf{Point} & $\boldsymbol{\Omega_r}$ & $\boldsymbol{\Omega_b}$ & $\boldsymbol{\Omega_c}$ & $\boldsymbol{\Omega_\phi}$ & $\boldsymbol{w_\phi}$ & \textbf{Existence Conditions}\\
\hline
A&0&0&0&1&-1&$\forall \lambda_\ast$\\
B&0&0&0&1&1&$\forall \lambda_\ast$\\
C&1&0&0&0&-&$\forall \lambda_\ast$\\
D$_c$&0&0&1&0&-&$\forall \lambda_\ast$ if $\beta=0$; never if $\beta\ne0$ \\
D$_b$&0&1&0&0&-&$\forall \lambda_\ast$\\
E&0&0&0&1&$-1+\frac{\lambda^2_\ast}{3}$&$\lambda_\ast^2<6$\\
F&0&$\forall \Omega_b$ if $\beta=0$; 0 if $\beta\ne0$&$\frac{\lambda_\ast^2-\beta\lambda_\ast-3}{(\lambda_\ast-\beta)^2}$&$\frac{\beta^2-\beta\lambda_\ast+3}{(\lambda_\ast-\beta)^2}$&$\frac{\beta(\lambda_\ast-\beta)}{3-\beta(\lambda_\ast-\beta)}$&$\lambda_\ast\neq\beta,\quad \beta^2-\beta\lambda_\ast \ge -\frac32,\quad \lambda_\ast^2-\beta\lambda_\ast \ge 3$\\
G&$1-\frac{4}{\lambda^2_\ast}$&0&0&$\frac{4}{\lambda^2_\ast}$&$\frac{1}{3}$&$\lambda_\ast^2\ge4$\\
\hline
\end{tabular}
}
\caption{Critical points of the constant-slope system.~In the uncoupled limit, these coincide with the corresponding fixed points of the full background dynamics.~In the coupled case, the matter-like point F survives as an exact critical point only in the baryon-free subsystem (i.e., for $\Omega_b=0$).~The same structure applies on both asymptotic branches of the double-exponential potential, upon identifying $\lambda_\ast=\lambda_{\rm early}$ or $\lambda_\ast=\lambda_{\rm late}$.}
\label{tab:fixed_points}
\end{table*}

We begin with the $\lambda_{\rm late}$ branch, whose role is twofold.~First, it must contain a viable late-time DE attractor.~Second, the trajectory must be able to approach this branch only after the early-time EDE episode has ended and the scalar contribution has been sufficiently depleted.~The natural candidate for the late-time endpoint is point $E$ in Table~\ref{tab:fixed_points}, which corresponds to a scalar-dominated solution capable of driving accelerated expansion, provided $\lambda_{\rm late}$ is sufficiently small.

The early-time branch with $\lambda=\lambda_{\rm early}$ is more constrained.~As discussed in Section~\ref{sec.3}, the evolution on this branch must originate from a radiation-dominated past in which the scalar field is subdominant.~This corresponds to the asymptotic state $\Omega_r \to 1$, $\Omega_c \to 0$, and $\Omega_\phi \to 0$, namely approach to the radiation-dominated point $C$.~The system must then enter an EDE phase.~The natural candidate is the radiation-scaling solution $G$, since it sustains a non-vanishing scalar fraction during radiation domination,
\begin{equation}
\Omega_\phi^{(G)} = \frac{4}{\lambda_{\rm early}^2}\,,
\end{equation}
with $w_\phi^{(G)}=1/3$.\footnote{A radiation-like phase with $w_\phi\simeq 1/3$ should be understood only as a statement about the homogeneous scalar-field equation of state.~In general, the perturbations of the scalar will not act as perfect fluid with $w=1/3$. Whether such a phase has an acceptable perturbation sector is therefore a separate, model-dependent question.}~The requirement $\Omega_\phi\simeq0.1$ therefore fixes $\lambda_{\rm early}^2\simeq40$.

After the EDE phase, the evolution proceeds toward epochs in which the matter sector becomes dynamically important.~The next relevant point is therefore $F$, the matter-era scaling solution.~The interpretation of point $F$ differs in the coupled and uncoupled cases.~In the uncoupled limit, baryons and cold dark matter both redshift as pressureless matter and are dynamically indistinguishable at the background level.~The matter-scaling solution therefore depends only on the total matter fraction, leaving $\Omega_b$ arbitrary.~When the coupling is switched on, this degeneracy is broken:~cold dark matter exchanges energy with the scalar, whereas baryons remain minimally coupled.~As a result, the scalar--cold-dark-matter scaling solution survives as an exact fixed point only on the $\Omega_b=0$ submanifold.~Thus, in the coupled case, point $F$ should be understood as a matter-like scaling solution of the dark sector rather than of the full baryon-plus-cold-dark-matter sector.

This distinction is important for determining whether the model can reproduce the dynamical requirements derived in Section~\ref{sec.3}.~We therefore treat the uncoupled and coupled cases separately, beginning with the simpler uncoupled case.

\subsubsection{Uncoupled case}

In the uncoupled case, point $F$ corresponds to a matter-dominated scaling regime in which the scalar behaves like pressureless matter, $w_\phi^{(F)}=0$, while retaining a fractional energy density
\begin{equation}
\Omega_\phi^{(F)} = \frac{3}{\lambda_{\rm early}^2}\simeq0.075\,,
\end{equation}
where we have imposed $\lambda_{\rm early}^2\simeq40$ as derived above.~This residual contribution is too large to satisfy the depletion and quiescent requirements derived in Section~\ref{sec.3}.~Indeed, since $\Omega_\phi^{(G)}=4/\lambda_{\rm early}^2$, one has
\begin{equation}
\frac{\Omega_\phi^{(F)}}{\Omega_\phi^{(G)}}=\frac{3}{4}\,,
\end{equation}
so the scalar is only mildly depleted after the EDE phase.~The uncoupled double-exponential model therefore fails to describe a viable unified EDE and late-time DE scenario:~after the EDE phase, the scalar settles onto a matter-like scaling solution rather than becoming sufficiently subdominant.

\begin{figure}[htpb!]
    \centering
    \includegraphics[width=0.8\linewidth]{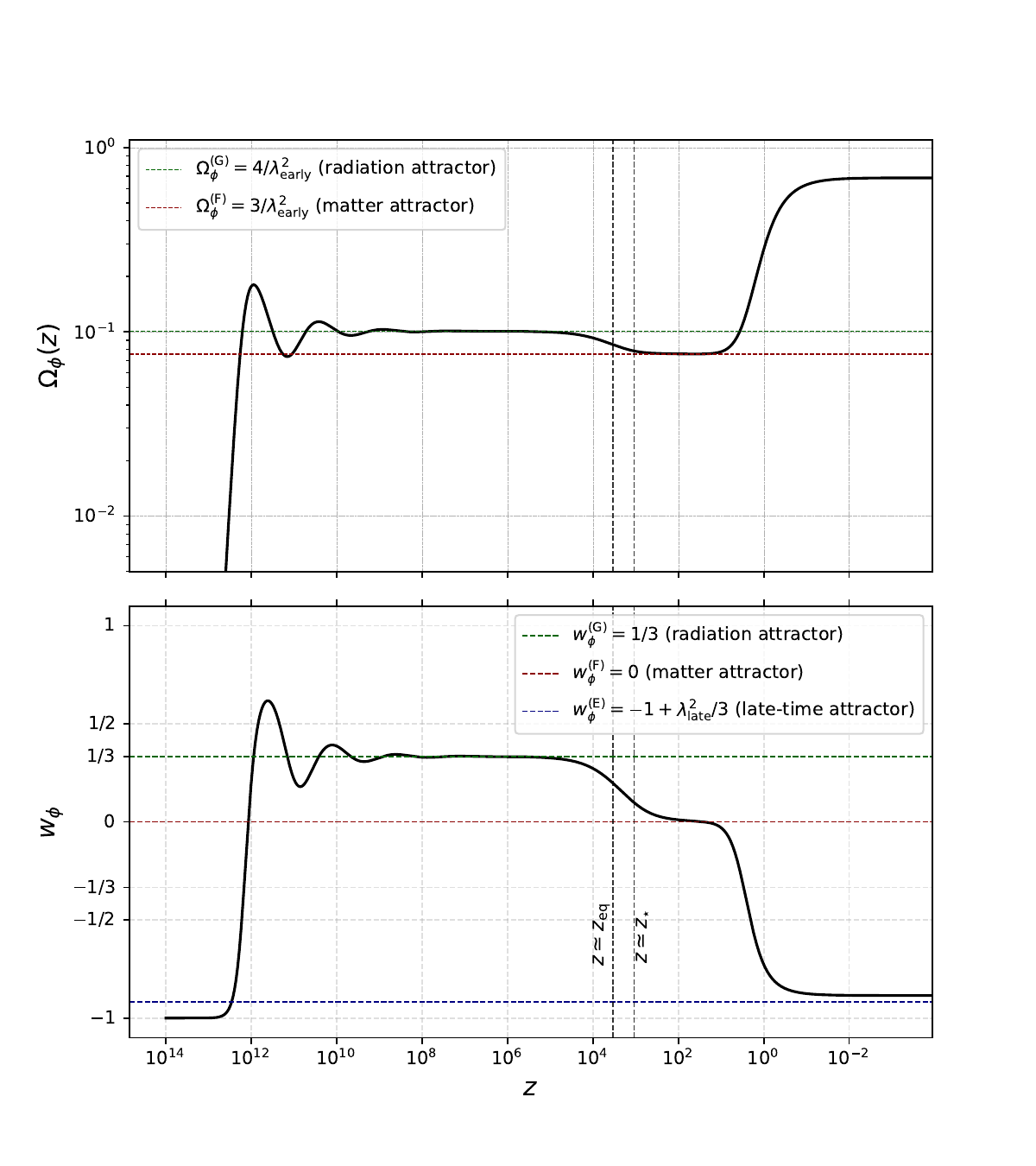}
    \caption{Representative trajectory of the uncoupled double-exponential model, obtained by solving the full cosmological evolution for $\lambda_{\rm early}^2 \simeq 40$ and $\lambda_{\rm late}^2 \simeq 0.25$.~The top panel shows the evolution of $\Omega_\phi(z)$, from the frozen regime to the radiation-scaling solution $G$ (green line), the subsequent matter-like scaling regime $F$ (red line), and the final late-time reactivation.~The bottom panel shows the corresponding evolution of $w_\phi(z)$, illustrating the transitions from the frozen phase to the radiation-like and matter-like tracking regimes, and finally toward the late-time attractor on the second branch (blue line).~The early-time amplitude of the potential is fixed so that the field starts rolling deep in the radiation-dominated era, while the late-time amplitude is fixed by shooting so that the present-day dark-energy budget is reproduced.}
    \label{fig:uncoupled_tracking}
\end{figure}

An example of the full background evolution, obtained {by implementing this model in a modified version of}~\texttt{CLASS}~\cite{Lesgourgues:2011re,Blas:2011rf}, is shown in Fig.~\ref{fig:uncoupled_tracking} with $\lambda_{\rm early}^2 \simeq 40$ and $\lambda_{\rm late}^2 \simeq 0.25$.~The representative trajectory follows the expected sequence $C\to G\to F\to E$, confirming that the fixed-point analysis captures the full background evolution.~Ultimately, the obstruction is that the same slope $\lambda_{\rm early}$ controls both the EDE amplitude and the subsequent matter-era tracking solution, so the depletion and quiescent requirements cannot be adjusted independently.

\subsubsection{Coupled case}
\label{sec.4.2.2}

\begin{figure}[ht!]
    \centering
    \includegraphics[width=0.8\linewidth]{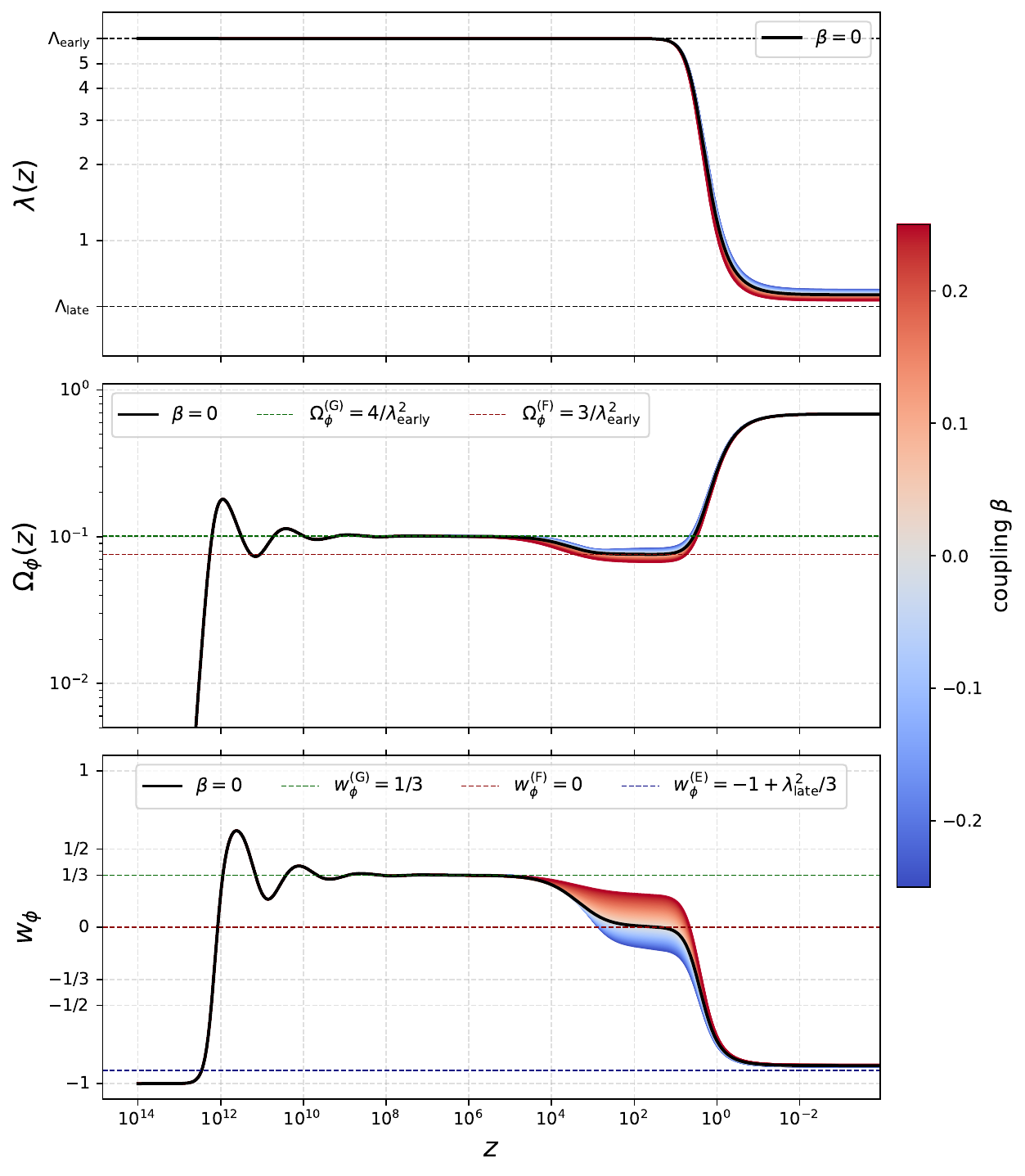}
    \caption{Representative trajectories of the coupled double-exponential model for different values of the conformal coupling, $\beta \in [-0.25,0.25]$, obtained by solving the full cosmological evolution for $\lambda_{\rm early}^2 \simeq 40$ and $\lambda_{\rm late}^2 \simeq 0.25$.~{The top panel shows the evolution of the logarithmic slope $\lambda(z)$.} The second panel shows evolution of $\Omega_\phi(z)$.~The bottom panel shows the corresponding evolution of $w_\phi(z)$.~The black solid curve denotes the uncoupled case, $\beta=0$, while the colored curves show the corresponding coupled evolutions.~The horizontal red and green lines mark the same uncoupled radiation- and matter-like reference solutions discussed in Fig.~\ref{fig:uncoupled_tracking}.~Small couplings remain very close to the uncoupled trajectory, while only much larger values produce substantial deformations of the intermediate evolution.}
\label{fig:coupled_tracking}
\end{figure}

As noted above, the minimally coupled case is obstructed by the persistence of the scaling solution into the matter era so that $75\%$ of the EDE energy density is retained.~The coupled case offers the potential to remove this obstruction because energy transfer from the scalar to cold dark matter can deplete its energy density during this phase.~Unfortunately, this is not possible, as we now demonstrate.

We start by considering the sub-system with $\Omega_b=0$. At the fixed point $F$, the scalar energy density is
\begin{equation} \label{eq:Omega_F_Coupled}
    \Omega_\phi^{(F)}
    =
    \frac{\beta^2-\beta\lambda_{\rm early}+3}
    {(\lambda_{\rm early}-\beta)^2}\,.
\end{equation}
However, this cannot be made arbitrarily small, because the fixed point only exists if
\begin{equation}
    \beta^2-\beta\lambda_{\rm early}+3\ge \frac{3}{2}
\end{equation}
(see Table~\ref{tab:fixed_points}).~The minimum residual scalar density is obtained when this bound is saturated, which occurs at
\begin{equation}\label{eq:beta_star}
    \beta_\star
    =
    \frac{1}{2}
    \left(
    \lambda_{\rm early}
    -
    \sqrt{\lambda_{\rm early}^2-6}
    \right)
    \simeq
    \frac{3}{2\lambda_{\rm early}}
\end{equation}
for $\lambda_{\rm early}\gg1$.~The corresponding minimum is
\begin{equation}
    \Omega_{\phi,\min}^{(F)}
    \simeq
    \frac{3}{2\lambda_{\rm early}^2}.
\end{equation}
Since the radiation-scaling point gives $\Omega_\phi^{(G)}={4}/{\lambda_{\rm early}^2}$, one finds, at best,
\begin{equation}
    \frac{\Omega_{\phi,\min}^{(F)}}{\Omega_\phi^{(G)}}
    \simeq
    \frac{3}{8}.
\end{equation}
Thus, for an EDE contribution $\Omega_\phi^{(G)}\simeq0.1$, the residual matter-era scalar fraction satisfies
\begin{equation}
    \Omega_\phi^{(F)}\gtrsim 0.0375,
\end{equation}
too large to satisfy the depletion and quiescent requirements derived above.~This conclusion is unchanged if the scalar is conformally coupled to baryons and cold dark matter with the same coupling $\beta$.~In that case, point $F$ becomes a fixed point for arbitrary $\Omega_b$, but the lower bound on the residual scalar fraction follows from the same existence condition and is therefore unaffected.

When $\Omega_b>0$, point $F$ ceases to be an exact fixed point of the full system. However, as shown in Appendix~\ref{app:baryon_linearized}, accounting for this does not evade the lower bound on the residual scalar fraction derived above. For $\beta>0$, baryons reduce the amount of dark matter coupled to the scalar, thereby weakening coupling-assisted depletion. For $\beta<0$, the coupled matter-era branch leaves a larger residual scalar fraction than the uncoupled scaling solution, so the coupling worsens the depletion problem rather than solving it.~We verified this by solving the background equations using \texttt{CLASS} with $\lambda_{\rm early}^2=40$ and $\beta\in[-0.25,0.25]$, the range where point $F$ exists at $\Omega_b=0$.~The evolution of $\Omega_\phi(z)$ and $w_\phi(z)$ is shown in Fig.~\ref{fig:coupled_tracking}.~Evidently, the fixed-point behavior above is qualitatively retained.

This can be understood from the evolution of $\lambda(z)$, shown in the top panel of Fig.~\ref{fig:coupled_tracking}.~For this potential, Eq.~\eqref{eq:lambdaprime_generic} can be written as  
\begin{equation}
\lambda'
=
-\sqrt{6}\,x\,(\lambda-\lambda_{\rm early})(\lambda-\lambda_{\rm late})\,,
\label{eq:lambda_prime_double_exp}
\end{equation}
which does not depend explicitly on $\beta$.~The asymptotic values $\lambda_{\rm early}$ and $\lambda_{\rm late}$ are fixed solely by the potential.~The conformal interaction can alter how quickly the trajectory moves between these branches, but it does not create a new constant-slope regime.

\subsection{General two-slope potentials}

The obstruction identified above is not specific to the double-exponential potential.~Rather, it reflects a more general limitation of potentials whose phase space contains only two asymptotic constant-slope regimes.~Whenever the trajectory evolves near one of these branches, the dynamics locally reduce to those of a single exponential potential.~Thus, if one branch is responsible for the EDE phase and the other for late-time acceleration, there is no independent constant-slope regime to generate the rapid depletion and quiescent phase required after EDE.~The only remaining loophole is that the transition from the early to the late branch might occur rapidly enough to prevent the trajectory from settling onto the matter-era scaling solution $F$.

To investigate this possibility, we consider the potential
\begin{equation}
V(\phi)=V_0\exp\left[
-\frac{\lambda_{\rm early}+\lambda_{\rm late}}{2}\phi
+\frac{1}{2}(\lambda_{\rm early}-\lambda_{\rm late})\Delta\phi\,
\log\cosh\left(\frac{\phi-\phi_\star}{\Delta\phi}\right)
\right],
\label{eq:V_step_slope}
\end{equation}
which has a rapid transition between two asymptotic slopes $\lambda_{\rm early}$ and $\lambda_{\rm late}$:
\begin{equation}
\lambda(\phi)=\frac{\lambda_{\rm early}+\lambda_{\rm late}}{2}
-\frac{\lambda_{\rm early}-\lambda_{\rm late}}{2}
\tanh\left(\frac{\phi-\phi_\star}{\Delta\phi}\right),
\label{eq:lambda_step_slope}
\end{equation}
where $\phi_\star$ is the field value where the transition occurs and $\Delta\phi$ sets the transition's width.~This potential captures the extreme limit where the two exponential regimes are realized at arbitrary field values.

There are two possible ways to try to evade the matter-era scaling point $F$.~First, one could choose $\phi_\star$ so that the transition from the early to the late slope occurs immediately after the EDE phase, before the trajectory settles onto $F$.~However, in this case the system is driven directly from the radiation-scaling regime $G$ toward the scalar-dominated point $E$ on the late branch.~This does not provide a viable depletion phase:~the late branch is shallow, with
\begin{equation}
w_\phi^{(E)}=-1+\frac{\lambda_{\rm late}^2}{3},
\end{equation}
so the scalar redshifts too slowly and instead begins behaving as late-time DE before its energy density has been sufficiently depleted.

The second possibility --- changing the sharpness of the transition --- does not remove the obstruction either.~A smaller $\Delta\phi$ makes the transition from $\lambda_{\rm early}$ to $\lambda_{\rm late}$ more abrupt, but it still sends the field directly from the EDE branch toward the shallow late-time branch.~It therefore does not provide a separate interval in which the scalar can redshift rapidly and become quiescent.~A larger $\Delta\phi$ only makes the interpolation smoother, causing the field to spend longer between the two branches, but again without generating an independent depletion regime.~Thus varying $\Delta\phi$ changes the details of the transition, not the underlying phase-space structure.~

Given the results above, we conclude that two-slope potentials cannot realize the required sequence of regimes for unified early- and late-time DE.~The simplest possible extension is to introduce a third slope, which we analyze next.

\section{Three-slope potentials}
\label{sec.5}

The analysis above provides guidance for constructing dynamically viable three-slope potentials.~The central obstruction is the matter-era scaling solution F.~If EDE is driven by a branch with $\lambda=\lambda_{\rm early}\simeq\sqrt{40}$, then the subsequent matter-era scaling solution leaves a residual scalar fraction
\begin{equation}
\Omega_\phi^{(F)}=\frac{3}{\lambda_{\rm early}^2}\simeq 0.075,
\end{equation}
which is too large to satisfy the depletion and quiescent requirements.~The minimal cure is therefore to introduce an intermediate, steeper branch after the EDE phase.~Requiring the residual scalar fraction on this branch to be at or below the percent level gives
\begin{equation}
\Omega_\phi^{(F)}=\frac{3}{\lambda_{\rm mid}^2}\lesssim 10^{-2},
\end{equation}
or equivalently
\begin{equation}
\lambda_{\rm mid}^2\gtrsim 300.
\end{equation}
Thus the potential must exhibit a steep--steeper--shallow hierarchy:~an initial moderately steep branch to generate EDE, a steeper intermediate branch to produce depletion and quiescence, and a shallow late-time branch to drive cosmic acceleration.

To proceed, we generalize the potential in \eqref{eq:V_step_slope} to the three-slope case as
\begin{equation}
\begin{aligned}
V(\phi)=V_0\exp\Bigg[
&-\frac{\lambda_{\rm early}+\lambda_{\rm late}}{2}\phi
+\frac{1}{2}(\lambda_{\rm early}-\lambda_{\rm mid})\Delta\phi_1
\log\cosh\left(\frac{\phi-\phi_{\star,1}}{\Delta\phi_1}\right)\\
&+\frac{1}{2}(\lambda_{\rm mid}-\lambda_{\rm late})\Delta\phi_2
\log\cosh\left(\frac{\phi-\phi_{\star,2}}{\Delta\phi_2}\right)
\Bigg],
\end{aligned}
\label{eq:V_double_step}
\end{equation}
with $\phi_{\star,1}<\phi_{\star,2}$.~This provides a minimal realization of the three-slope structure required by the dynamical arguments above, serving as a controlled prototype for the more realistic model-building possibilities discussed below:
\begin{equation}
\lambda(\phi)=\frac{\lambda_{\rm early}+\lambda_{\rm late}}{2}
-\frac{\lambda_{\rm early}-\lambda_{\rm mid}}{2}
\tanh\left(\frac{\phi-\phi_{\star,1}}{\Delta\phi_1}\right)
-\frac{\lambda_{\rm mid}-\lambda_{\rm late}}{2}
\tanh\left(\frac{\phi-\phi_{\star,2}}{\Delta\phi_2}\right).
\label{eq:lambda_double_step}
\end{equation}
By construction, $\lambda(\phi)$ approaches $\lambda_{\rm early}$ for $\phi\ll\phi_{\star,1}$, $\lambda_{\rm mid}$ for $\phi_{\star,1}\ll\phi\ll\phi_{\star,2}$, and $\lambda_{\rm late}$ for $\phi\gg\phi_{\star,2}$.

\begin{figure}[tpb!]
    \centering
    \includegraphics[width=0.8\linewidth]{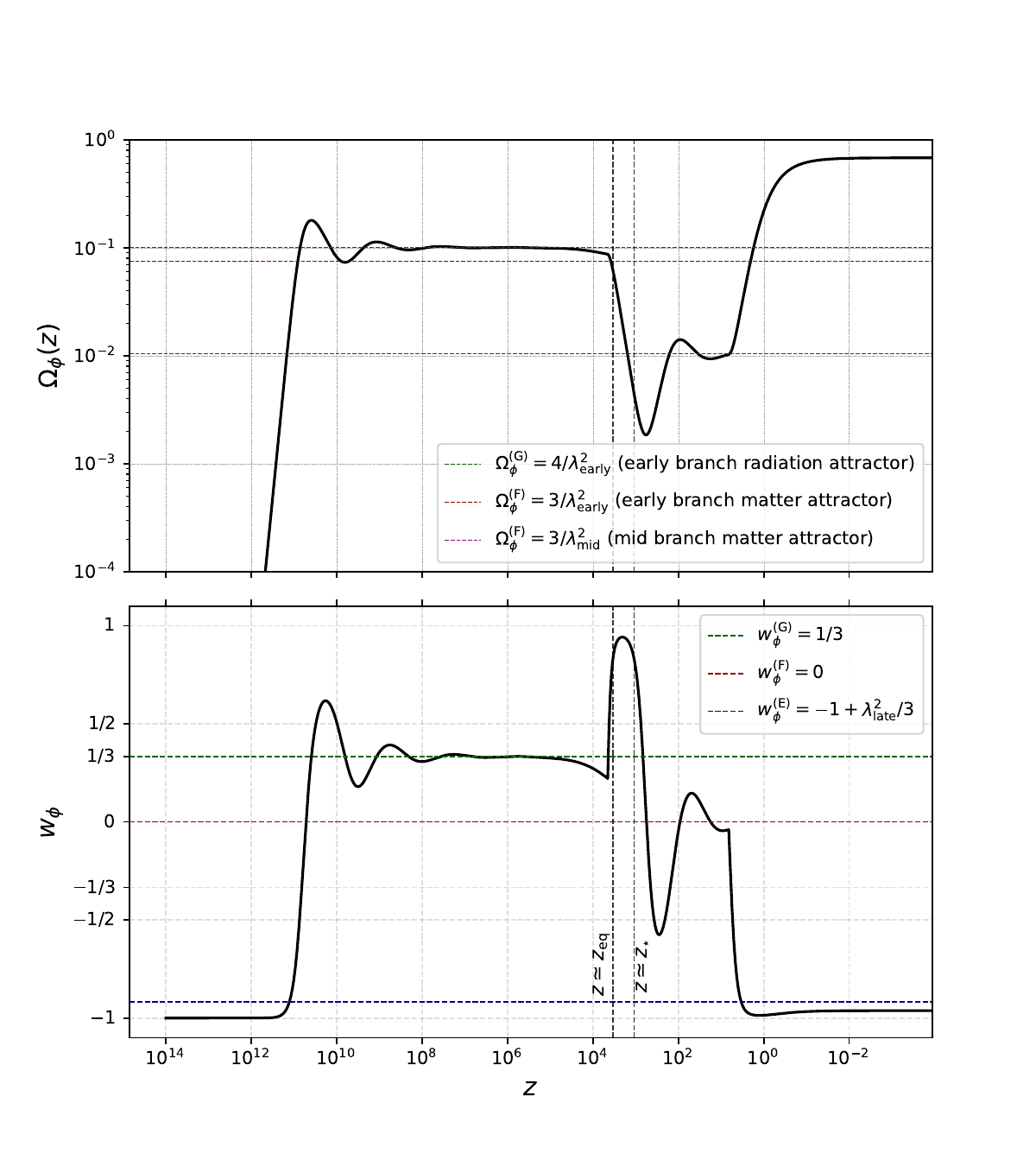}
    \caption{Representative trajectory of the three-slope potential in Eq.~\eqref{eq:V_double_step}.~The top panel shows the evolution of $\Omega_\phi(z)$, and the bottom panel shows the corresponding evolution of $w_\phi(z)$.~The field first follows the early branch and reaches the radiation-scaling regime associated with the EDE phase.~The trajectory is subsequently redirected toward the much steeper intermediate branch, whose matter-like scaling level, shown by the purple horizontal line, lies well below the corresponding early-branch value, shown by the red horizontal line.~This produces efficient suppression of the scalar abundance and a prolonged quiescent phase before the system finally evolves toward the shallow late-time branch responsible for DE.~Although the trajectory does not fully settle onto the intermediate asymptotic fixed point before the late-time transition begins, its attraction toward that branch is sufficient to realize the required depletion.}
    \label{fig:3H_tracking}
\end{figure}

In Fig.~\ref{fig:3H_tracking} we show a representative trajectory obtained by implementing this potential in \texttt{CLASS}.~We take the same early and late slopes as in the two-slope examples, $\lambda_{\rm early}^2\simeq40$ and $\lambda_{\rm late}^2\simeq0.25$, and set the intermediate steeper slope to $\lambda_{\rm mid}^2\sim300$.~The resulting evolution realizes the full sequence identified in Section~\ref{sec.3}:~an initial frozen phase, an EDE-like radiation-scaling phase, efficient depletion, a prolonged quiescent epoch, and late-time reactivation responsible for the present accelerated expansion.

The mechanism is exactly the one anticipated from the fixed-point analysis.~The EDE amplitude is still controlled by the early branch, with $\Omega_\phi^{(G)}\sim4/\lambda_{\rm early}^2\sim0.1$.~After the EDE phase, however, the trajectory is redirected onto the steeper intermediate branch rather than remaining on the early branch and approaching the matter-era scaling point with $\Omega_\phi^{(F)}\sim3/\lambda_{\rm early}^2$.~The residual scalar fraction is instead controlled by
\begin{equation}
\Omega_\phi^{(F,{\rm mid})}\sim \frac{3}{\lambda_{\rm mid}^2}\,,
\end{equation}
which is of order $10^{-2}$.~This provides the required depletion and leaves a sufficiently small post-recombination scalar fraction for the quiescent phase.

Note that the trajectory need not settle exactly onto the asymptotic fixed point of the intermediate branch.~What matters is that the post-EDE evolution is redirected toward a sufficiently steep region of the potential before the trajectory relaxes onto the matter-like plateau associated with the early branch.~The later transition to the shallow branch then allows the same field to approach the late-time DE attractor.~

\subsection{Model-building implications}

\begin{figure}[t!]
    \centering
    \includegraphics[width=0.75\linewidth]{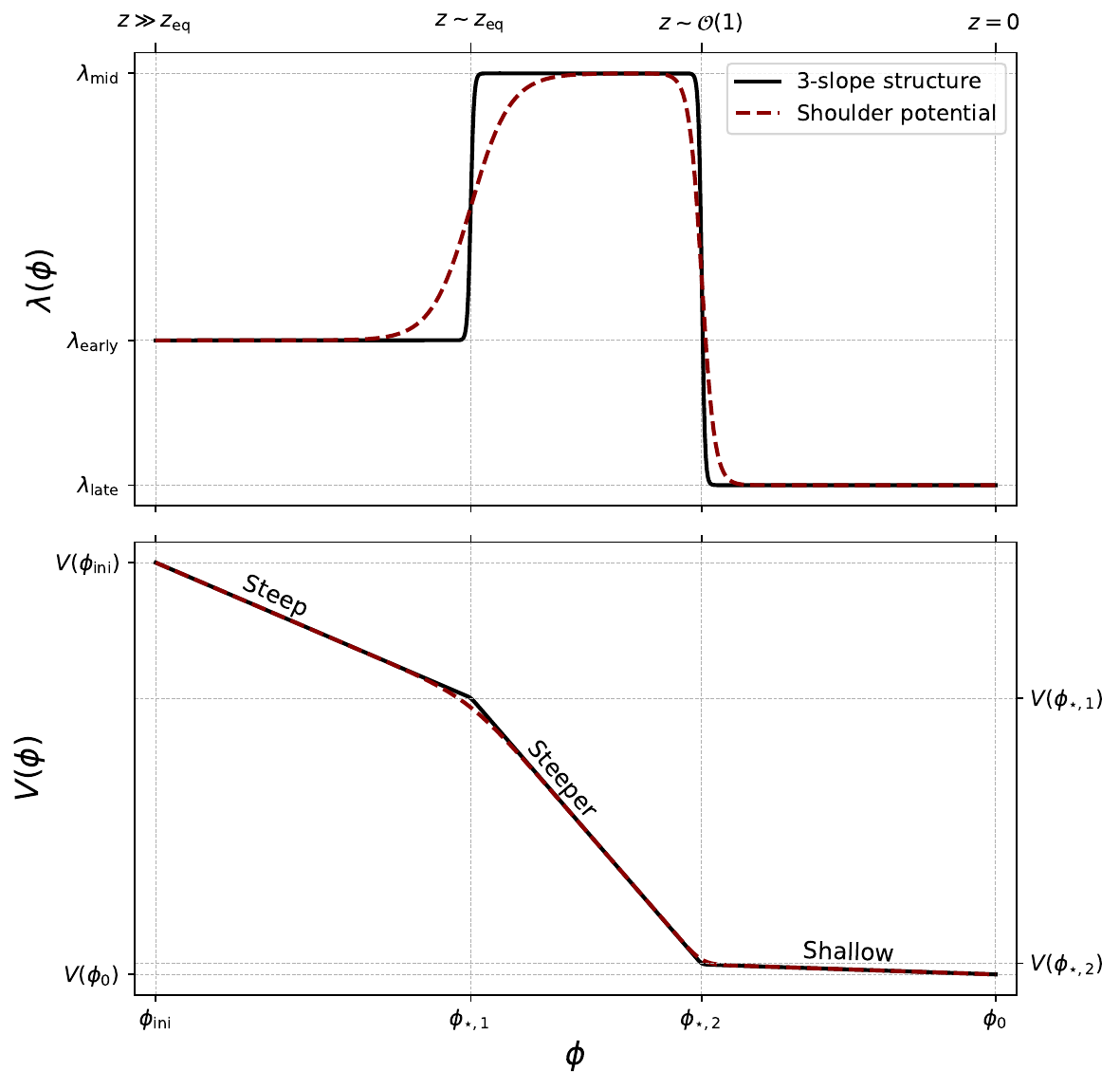}
    \caption{Schematic representation of the minimal three-hypersurface phase-space geometry required to realize the full sequence of frozen evolution, EDE, depletion, quiescence, and late-time reactivation. The black curves show the required three-slope geometry parametrized by the potential in Eq.~\eqref{eq:V_double_step}, while the dark-red dashed curves show an explicit realization based on the shoulder-like potential in Eq.~\eqref{eq:V_three_slope_exp}. The top panel shows the non-monotonic logarithmic slope $\lambda(\phi)$, which interpolates between an early steep branch $\lambda_{\rm early}$, an intermediate steeper branch $\lambda_{\rm mid}$, and a shallow late-time branch $\lambda_{\rm late}$. The upper horizontal axis indicates the qualitative cosmological epochs at which the transitions are required to occur along the trajectory. The bottom panel shows the corresponding potential profiles $V(\phi)$ in logarithmic scale. The three branches correspond respectively to the EDE regime, the depletion and quiescent regime, and the late-time DE regime.}
    \label{fig:3H_summary}
\end{figure}

Having identified the minimal set of features required for a potential to provide a unified description of EDE and late-time DE, we now turn to the question of which concrete potentials can realize this structure.~The necessary features are summarized in~Fig.~\ref{fig:3H_summary}.~The top panel shows $\lambda(\phi)$, which interpolates between the three asymptotic constant-slope hypersurfaces $\lambda_{\rm early}$, $\lambda_{\rm mid}$, and $\lambda_{\rm late}$.~The bottom panel shows the corresponding potential $V(\phi)$ using a logarithmic scale.~The steep--steeper--shallow transition is apparent.

Realizing the required sequence of slope transitions is non-trivial.~A natural first attempt is a triple-exponential potential,
\begin{equation}
    V(\phi)
    =
    V_{\rm early}e^{-\lambda_{\rm early}\phi}
    +
    V_{\rm mid}e^{-\lambda_{\rm mid}\phi}
    +
    V_{\rm late}e^{-\lambda_{\rm late}\phi},
\end{equation}
with
\begin{equation}
    \lambda_{\rm mid}\gg \lambda_{\rm early}\gg \lambda_{\rm late}.
\end{equation}
However, this does not generically realize the desired sequence of slopes.~For the usual ordering of positive slopes and monotonic field evolution, the steepest term dominates at early field values, so the trajectory encounters the branches in the order
\begin{equation}
    \lambda_{\rm mid}\to \lambda_{\rm early}\to \lambda_{\rm late},
\end{equation}
rather than the required
\begin{equation}
    \lambda_{\rm early}\to \lambda_{\rm mid}\to \lambda_{\rm late}.
\end{equation}

This obstruction is generic for positive sums of exponentials.~Consider
\begin{equation}
V(\phi)=\sum_{i=1}^{N} V_i e^{-\lambda_i \phi},
\qquad V_i>0 .
\label{eq:sum_exp}
\end{equation}
The logarithmic slope is
\begin{equation}
\lambda(\phi)\equiv -\frac{V_{,\phi}}{V}
=
\sum_i w_i(\phi)\lambda_i,
\qquad
w_i(\phi)=
\frac{V_i e^{-\lambda_i\phi}}
{\sum_j V_j e^{-\lambda_j\phi}},
\end{equation}
with $w_i>0$ and $\sum_i w_i=1$.~Thus $\lambda(\phi)$ is a convex weighted average of the slopes.~Differentiating gives
\begin{equation}
\frac{d\lambda}{d\phi}
=
-\left[
\sum_i w_i\lambda_i^2
-
\left(\sum_i w_i\lambda_i\right)^2
\right]
=
-{\rm Var}_w(\lambda_i)
\le 0.
\end{equation}
Therefore, for any positive sum of exponentials, $\lambda(\phi)$ is monotonically non-increasing as the field rolls toward larger $\phi$.~Such potentials can interpolate only from larger to smaller slopes; they cannot realize the required sequence $\lambda_{\rm early}\to\lambda_{\rm mid}\to\lambda_{\rm late}$ with $\lambda_{\rm mid}>\lambda_{\rm early}>\lambda_{\rm late}$.

Beyond exponentials, our results suggest that viable potentials must exhibit a localized slope-steepening feature to realize the depletion and quiescent phases, e.g., a bump or shoulder.~As an example, consider the potential
\begin{equation}
    \label{eq:V_three_slope_exp}
    V(\phi)=V_{\rm late}e^{-\lambda_{\rm late}\phi}+V_{\rm early}\frac{e^{-\lambda_{\rm early}\phi}}{\left(1+\exp\left[\frac{\phi-\phi_\star}{\Delta}\right]\right)^n}\,,\qquad\textrm{with}\qquad\lambda_{\rm early}\gg\lambda_{\rm late}\,,
\end{equation}
which behaves as a double exponential with a shoulder feature at $\phi=\phi_\star$ with width $\Delta$.~When $\phi\ll\phi_\star$ this behaves as a double exponential of the form
\begin{equation}
    \label{eq:V_three_slope_exp_early}
     V(\phi)=V_{\rm late}e^{-\lambda_{\rm late}\phi}+V_{\rm early}{e^{-\lambda_{\rm early}\phi}}\simeq V_{\rm early}{e^{-\lambda_{\rm early}\phi}}\,,
\end{equation}
so, given the hierarchy, this is effectively a single exponential with $\lambda=\lambda_{\rm early}$.~This provides the early branch responsible for the frozen and EDE phases.~When $\phi\gg\phi_\star$ the potential also behaves as a double exponential but of the form
\begin{equation}
    \label{eq:V_three_slope_exp_late}
     V(\phi)=V_{\rm mid}{e^{-\lambda_{\rm mid}\phi}+}V_{\rm late}e^{-\lambda_{\rm late}\phi},
\end{equation}
with 
\begin{equation}
    \lambda_{\rm mid}=\lambda_{\rm early}+\frac{n}{\Delta}\qquad\textrm{and}\qquad V_{\rm mid}=V_{\rm early}e^{n\frac{\phi_\star}{\Delta}}.
\end{equation}
This double exponential then includes the steeper slope needed to account for the depletion and quiescent phase and the shallow slope needed for realizing late DE.~Taking the benchmarks above, $\lambda_{\rm early}^2=40$ and $\lambda_{\rm mid}^2=300$, implies
\begin{equation}
    \frac{n}{\Delta}\approx11.
\end{equation}
Thus a modest value such as $n=2$ with $\Delta\simeq0.18$ already produces the required intermediate steepening. This explicit realization is shown by the dark-red dashed curves in Fig.~\ref{fig:3H_summary}, where it is compared with the required three-slope geometry.~While this potential is a useful phenomenological realization of the required structure, it should not be interpreted as a UV-derived model. Rather, it illustrates the model-building challenge:~a fundamental construction must generate a localized shoulder capable of producing a sufficiently large temporary steepening of the logarithmic slope.

\section{Discussion and conclusions}
\label{sec.6}

In this work, we have investigated the possibility that early dark energy, a proposed resolution of the Hubble tension, and late dark energy, could be manifestations of a single scalar field.~

We first derived a set of model-independent requirements that a scalar must satisfy to produce a viable phenomenology.~We found that the scalar must experience five distinct phases, summarized in Fig.~\ref{fig:target_EDE}:
\begin{itemize}
    \item \textit{\textbf{Frozen phase:}} at very early times the field must remain Hubble-damped and subdominant, with $\Omega_\phi \ll \Omega_r \simeq 1$ and $w_\phi \simeq -1$, so as not to interfere with the standard radiation-dominated evolution.
    
    \item \textit{\textbf{EDE phase:}} the field must be 
    active sometime after $z\sim{\rm few}\times10^4$ and before matter-radiation equality.~During this time, it must constitute $\simeq10\%$ of the universe's energy budget in order to reduce the sound horizon sufficiently to resolve the Hubble tension.~
        
    \item \textit{\textbf{Depletion phase:}} to preserve consistency with the CMB, the scalar's contribution to the energy budget must rapidly deplete  after matter-radiation equality, reaching 
    percent levels {or below} by recombination.
    
    \item \textit{\textbf{Quiescence:}} after depletion, the field must remain  subdominant throughout the post-recombination history, so as not to induce an appreciable correction to the comoving distance to last scattering.~This is necessary to preserve the geometric mechanism through which the reduction of the sound horizon is translated into a higher inferred value of $H_0$.
    
    \item \textit{\textbf{Late-time reactivation:}} finally, the same degree of freedom must begin to dominate the cosmological dynamics at sufficiently low redshift, so as to reproduce a late-time DE phase.
\end{itemize}

Next, in order to capture a large class of models without fine-tuning we attempted to realize each phase as a fixed point of the scalar's dynamics.~We found that the most natural potentials with two slopes e.g., double exponential, are unable to accommodate all of the requirements due to the presence of a scaling fixed point in the matter-era where the scalar contributes $7.5\%$ of the energy budget, in tension with the depletion and quiescent requirements.~Coupling the scalar to dark matter can reduce this fraction to at best $3.75\%$, which is still too high for a viable cosmology.~

We subsequently studied three-slope potentials, finding that they can provide a unified description of early- and late-time dark energy provided that they exhibit a steep--steeper--shallow structure; see Fig.~\ref{fig:3H_summary}.~The simplest monotonic potentials are unable to realize this hierarchy.~Instead, viable models require a localized steepening of the logarithmic slope, corresponding to a shoulder- or cliff-like feature in the potential, which may pose non-trivial model-building challenges.

Looking ahead, our results suggest several directions for future investigations.~First, one could study other classes of scalar field theories such as non canonical kinetic terms e.g., K-essence, Horndeski and beyond to derive similar requirements for the free functions appearing in the action.~Similarly, one could consider more general couplings to (dark) matter e.g., disformal couplings.~A second complementary direction would be to find string theory or supergravity constructions that can realize the necessary steep--steeper--shallow potential structure illustrated by Eq.~\eqref{eq:V_three_slope_exp}.

Finally, as with any no-go-type obstruction, it is important to understand its domain of validity.~Our conclusion assumes that the EDE component is depleted by the smooth rolling dynamics of the same scalar field that later sources dark energy, and that the scalar evolution can be described by successive quasi-autonomous phases with approximately constant logarithmic slope.~Additional dynamical processes could evade this requirement.~For example, a phase transition, decay into another sector, particle-production event, or other dissipative mechanism could remove the scalar energy density after the EDE phase without requiring an intermediate steep branch in the potential.~Similarly, potentials with rapid, non-adiabatic variations of the logarithmic slope may evade the local fixed-point description used in our dynamical-systems analysis. Field-dependent conformal couplings $\beta(\phi)$ may also modify the fixed-point structure and could evade some of the restrictions derived here for constant $\beta$.~These possibilities lie outside the smooth single-field tracking framework studied here and should be analyzed separately.

\section*{Acknowledgments}
W.G.~and J.S.~acknowledge support from National Aeronautics and Space Administration (NASA) under Grant No.~80NSSC24K0898.~

\appendix

\section{Effect of baryons on the coupled matter-era dynamics}
\label{app:baryon_linearized}

In Section~\ref{sec.4.2.2} we showed that, in the subsystem with $\lambda=\lambda_{\rm early}$ and $\Omega_b=0$, the coupled matter-era scaling solution, point $F$, exists only if
\begin{equation}
    \beta^2-\beta\lambda_{\rm early}+3\geq \frac{3}{2}.
\end{equation}
This condition implies a lower bound on the residual scalar fraction,
\begin{equation}
    \Omega_{\phi,\min}^{(F)}
    \simeq
    \frac{3}{2\lambda_{\rm early}^2},
    \qquad
    \frac{\Omega_{\phi,\min}^{(F)}}{\Omega_\phi^{(G)}}
    \simeq
    \frac{3}{8},
    \label{eq:coupled_depletion_ratio_appendix}
\end{equation}
where $\Omega_\phi^{(G)}=4/\lambda_{\rm early}^2$ is the radiation-scaling value. Thus, in the baryon-free subsystem, the coupled branch leaves an order-one fraction of the EDE abundance.~In practice, $\Omega_b>0$ at all times, so it is important to determine whether this invalidates our conclusions.~In this appendix, we show that it does not.

We begin by noting that, near fixed point $F$, the evolution of $\Omega_b$ is decoupled from the dynamics in the $x$-$y$ sub-manifold.~For $\Omega_r=0$, the baryon perturbation has eigenvalue \begin{equation}
    \mu_b = 3(x_F^2-y_F^2)
    = \frac{3\beta}{\lambda_{\rm early}-\beta}
    \simeq 3\frac{\beta}{\lambda_{\rm early}},
\end{equation}
for $\beta\ll\lambda_{\rm early}$.~This implies that $\Omega_b$ changes
only over an e-fold interval
\begin{equation}
    \Delta N_b \sim \frac{\lambda_{\rm early}}{3\beta}\gg 1.
\end{equation}
By contrast, the eigenvalues of the $x$--$y$ subsystem satisfy
\begin{equation}
    \mathrm{Re}(\mu_\pm)
    =
    -\frac{3}{4}\left(1-\frac{\beta}{\lambda_{\rm early}}\right),
\end{equation}
so $x$ and $y$, and hence $\Omega_\phi$, relax over an $\mathcal{O}(1)$
number of e-folds.~This implies that we can treat the system as quasi-static with $\Omega_b$ slowly evolving as
\begin{equation}
    \Omega_b(N)
    =
    \Omega_b(N_i)
    \exp\left[
    \frac{3\beta}{\lambda_{\rm early}}
    (N-N_i)
    \right].
    \label{eq:Omega_b_growth_appendix}
\end{equation}

Writing 
\begin{equation}
    x=x_F+\delta x\qquad\textrm{and}\qquad y=y_F+\delta y,
\end{equation}
we can linearize equations~\eqref{eq:xprime_generic} and \eqref{eq:yprime_generic}, taking $\lambda=\lambda_{\rm early}$ and $\beta\ll\lambda_{\rm early}$, to find
\begin{equation}
    \delta x=\sqrt{\frac{3}{2}}\beta\frac{\Omega_b}{3-\lambda_{\rm early} ^2}\qquad\textrm{and}\qquad\delta y =-\beta\frac{\Omega_b}{\sqrt{6-4 \beta  \lambda_{\rm early} }}.
\end{equation}
Using these expressions in Eq.~\eqref{eq:wphi_xy}, one finds that the energy density deviates from its value at fixed point $F$ in Eq.~\eqref{eq:Omega_F_Coupled} by an amount
\begin{equation}
    \Delta\Omega^{(F)}_\phi=-\frac{\beta}{\lambda_{\rm early}}  \left(1-\frac{3}{\lambda_{\rm early} ^2-3}\right)\Omega_b.
\end{equation}
To estimate the size of this shift, we take $\beta=\beta_\star$, corresponding to the minimum residual energy density, see Eq.~\eqref{eq:beta_star}, and $\lambda_{\rm early}^2=40$. This gives $\Delta\Omega^{(F)}_\phi\simeq0.034\Omega_b\simeq2\times10^{-3}$ for $\Omega_b\simeq0.05$.

For general $\beta$, we begin by noting that Equation \eqref{eq:Omega_b_growth_appendix} implies that $\Omega_b$ evolves towards the fixed point $F$ for $\beta<0$ but is repelled from it for $\beta>0$.~In the former case, the residual scalar energy density is larger than the uncoupled case (recall $\lambda_{\rm early}^2\simeq40$ for a sufficient EDE phase):
\begin{equation}
    \Omega_\phi^{(F)}-\frac{3}{\lambda_{\rm early}^2}
    \simeq - \frac{\beta}{\lambda_{\rm early}}\left(1-\frac{6}{\lambda_{\rm early}^2}\right),
\end{equation}
which is positive for $\beta<0$.~For $\beta>0$, we can perturb the Friedmann constraint to find
\begin{equation}
    \delta\Omega_c=-\Omega_b-\Delta\Omega^{(F)}_\phi=-\left[1-\frac{\beta}{\lambda_{\rm early}}  \left(1-\frac{3}{\lambda_{\rm early} ^2-3}\right)\right]\Omega_b.
\end{equation}
Thus $\Omega_c$ is driven to smaller values, reducing the effective forcing term for $x$,  $x' \supset \sqrt{\frac{3}{2}}\,\beta\,\Omega_c $ in Equation~\eqref{eq:xprime_generic}, so that the scalar depletion is diminished.~These results are consistent with Fig.~\ref{fig:coupled_tracking}.

\bibliographystyle{JHEP}

\bibliography{bib}

\end{document}